\documentclass[aps,prc,preprint,superscriptaddress,showpacs]{revtex4}

\newcommand{\beq}{\begin{equation}}
\newcommand{\eeq}{\end{equation}}
\newcommand{\beqa}{\begin{eqnarray}}
\newcommand{\eeqa}{\end{eqnarray}}

\newcommand{\bma}{\begin{mathematica}}
\newcommand{\ema}{\end{mathematica}}

\newcommand{\vk}{\mbox {\boldmath $k$\unboldmath}}
\newcommand{\vq}{\mbox {\boldmath $q$\unboldmath}}

\def\nn{\nonumber}

\usepackage{epsfig}

\begin{document}
\title{Dynamical coupled-channel approach to hadronic and electromagnetic
  production of kaon-hyperon on the proton
}

\author{B. Juli\'a-D\'{\i}az}
\email[]{bjulia@ecm.ub.es} \affiliation{Laboratoire de recherche
sur les lois fondamentales de l'Univers, DSM/Dapnia, CEA/Saclay,
91191 Gif-sur-Yvette, France}

\author{B. Saghai}
\email[]{bijan.saghai@cea.fr} \affiliation{Laboratoire de
recherche sur les lois fondamentales de l'Univers, DSM/Dapnia,
CEA/Saclay, 91191 Gif-sur-Yvette, France}

\author{T.-S. H. Lee}
\email[]{lee@phy.anl.gov}
\affiliation{Physics Division, Argonne National Laboratory,
Argonne, IL 60439, USA}

\author{F. Tabakin}
\email[]{tabakin@pitt.edu}
\affiliation{Department of Physics and Astronomy, University of Pittsburgh, USA}
\date{\today}

\begin{abstract}
A dynamical coupled-channel formalism for processes $\pi N~\to~KY$
and $\gamma N~\to~KY$ is presented which provides a comprehensive
investigation of recent data on the $\gamma p \to K^+ \Lambda$
reaction. The non-resonant interactions within the subspace
$KY\oplus\pi N$ are derived from effective Lagrangians, using a
unitary transformation method. The calculations of photoproduction
amplitudes are simplified by casting the coupled-channel equations
into a form such that the empirical $\gamma N \to \pi N$
amplitudes are input and only the parameters associated with the
$KY$ channel are determined by performing $\chi^2$-fits to all of
the available data for $\pi^- p \to
K^\circ\Lambda,~K^\circ\Sigma^\circ$ and $\gamma p \to
K^+\Lambda$. Good agreement between our models and those data are
obtained. In the fits to  $\pi N~\to~KY$ channels, most of the
parameters are constrained within $\pm 20\%$ of the values given
by the Particle Data Group and/or quark model predictions, while
for $\gamma p \to K^+ \Lambda$ parameters, ranges compatible with
 broken $SU(6)\otimes O(3)$ symmetry are imposed. The main
reaction mechanisms in $K^+ \Lambda$ photoproduction are singled
out and issues related to newly suggested resonances $S_{11}$,
$P_{13}$, and $D_{13}$ are studied. Results illustrating the
importance of using a coupled-channel treatment are reported.
Meson cloud effects on the $\gamma N \to N^*$ transitions are also
discussed.

\end{abstract}

\maketitle

\section{Introduction}
\label{sec:i}

Recent experiments at JLab-CLAS~\cite{JLab05,JLab04},
ELSA-SAPHIR~\cite{ELSA,SAPHIR05} and
Spring-8-LEPS~\cite{LEPS,LEPS05} are refining our knowledge of
associated strangeness photoproduction. High precision
differential cross section data for the process $\gamma p \to K^+
\Lambda$ have been released~\cite{JLab05,ELSA,LEPS05} covering the
region between $W~\approx$ 1.6 GeV and 2.6 GeV in the
center-of-mass frame. Furthermore, single polarization asymmetry
data for recoil hyperon~\cite{JLab04} and beam~\cite{LEPS,LEPS05}
have also become available.

The $K^+ \Lambda$ photoproduction has also been extensively
studied using phenomenological approaches. In general, those
works~\cite{ELA-SL,ELA-MB,ELA-SK,ELA-BE,Ireland04,Byd04}
investigated the direct channel mechanisms based on an isobar
approach in tree approximation. Combinations of isobar models with
a Regge analysis~\cite{REG-MLV}, successful at higher
energies, have also focused~\cite{REG-MB,REG-Gent} on strangeness
electromagnetic production. A new generation of more precise data
has made it clear that coupled-channel effects can no longer be
ignored and that multi-step processes have to be incorporated
carefully. Coupled-channel formalisms based on the K-matrix
approximation and isobar effective Lagrangians have been
developed~\cite{K-matrix-1, K-matrix-2}.

The purpose of this work is to report on an advanced version of a
dynamical coupled-channel formalism~\cite{CC-01,CC-04,CC-05a,CC-05b} that
incorporates proper treatment of off-shell effects. The direct
$KY$ photoproduction channel is investigated via a chiral
constituent quark model (CQM)~\cite{CQM-1a,CQM-1b}. This latter
approach allows one to handle all known resonances with a
reasonable number of adjustable parameters, in contrast to isobar
effective Lagrangian models~\cite{CQM-2}. Consequently, the CQM
provides an appropriate tool for understanding the elementary
reaction mechanism, establishing reliable indicia for the
predicted missing baryon
resonances~\cite{Isgur,Cap86,Cap92,Cap94,Cap98,CR2000,Bijker,Gian},
and gaining improved insights into the known resonances.

In principle, the $KY$ photoproduction should be investigated
within a large scale coupled-channel approach including several
reaction channels, e.g. $\pi N$, $\eta N$, $\omega N$, $KY$, $\phi
N$, $\pi\pi N$ ($\sigma N,~\pi\Delta,~\rho N$). Obviously,
this can not be done so easily because the data sets, to
be simultaneously fitted, are very extensive, and reaction
mechanisms involving channels other than $\pi N$ have not been
studied extensively.

As a first significant step, it is useful to consider a much more
restricted coupled-channel model focusing on understanding
particular reaction mechanisms. Concerning the $KY$
photoproduction, the obvious first task is to investigate the
coupling between the $KY$ and $\pi N$ channels for the following
reasons. From the available data, one observes that kaon
photoproduction is in general much weaker than pion
photoproduction. Hence the multi-step transitions, such as $\gamma
N \to \pi N \to KY$, should be comparable to the direct $\gamma N
\to KY$ process. This has been verified in Ref.~\cite{CC-01} using
a coupled-channel model with $\gamma N$, $\pi N$ and $KY$
channels. Moreover, the need for a coupled-channel approach to
study meson-baryon reactions in the second and third $N^*$ regions
has been well discussed in the literatures, as reviewed in
Refs.~\cite{lms,bl}.

In this work, we take advantage of the development of new
models~\cite{CC-04} for $\pi N \to KY$ and $KY\to KY$ interactions
to reinvestigate the influence of the $\pi N$ channel.
Furthermore, we refine the models developed in
Refs.~\cite{CC-01,CC-04} and consider recent $\gamma p \to
K^+\Lambda$ data. Focusing on the coupled-channel effects
associated with the $\pi N$ channel, we also determine the
parameters of relevant $N^*$ resonances. Our results could serve
as the starting point for performing more advanced coupled-channel
calculations including additional meson-baryon channels.

Within the considered coupled-channel model, a comprehensive study
of $K^+\Lambda$ photoproduction requires models of the
non-resonant transitions among $\gamma N$, $\pi N$, and $KY$
states and the decays into these three channels for about 12
isospin $I=1/2$ $N^*$ states. In this work, we follow
Refs.~\cite{CC-01,CC-04} to derive the non-resonant transitions
from effective Lagrangians by using a unitary transformation
method~\cite{SatoLee} and SU(3) symmetry. For $N^*$ decays, we
consider information from the Particle Data Group~\cite{PDG} (PDG)
and/or from constituent quark model
predictions~\cite{Cap92,Cap94,Cap98}. With these constraints, the
model has a reasonable number of adjustable parameters, which can
only be ascertained  from the data. We simplify the fitting task
by casting the coupled-channel equations into a form such that the
empirical $\gamma N \to \pi N$ amplitudes~\cite{SAID} are input to
the calculations and only the parameters associated with the $KY$
channel are to be determined by performing $\chi^2$ minimization
fits to all available  $\pi N \to KY$ and $\gamma N \to KY$ data,
using the CERN-MINUIT code.

In addition, to clarify the role of coupled-channel effects due to
the $\pi N$ channel, we also analyze the dynamical content of the
$\gamma N \to N^*$ transition. The so-called ``meson cloud
effects'' discussed in the study~\cite{SatoLee,ky} of the $\Delta$
(1232) resonance are identified within our coupled-channel model.
We also make an attempt to determine the properties of the
predicted~\cite{Cap86,Cap92,Cap94,Cap98,CR2000,Bijker,Gian,LW96}
and/or sought 
for~\cite{ELA-MB,Ireland04,REG-MB,K-matrix-1,CC-05a,CQM-1b,
LS-3,Zagreb,WTC,chen,Try04,BES-1,BES-2,Kelkar} third $S_{11}$,
$P_{13}$, and $D_{13}$ resonances.

For simplicity, at this stage we do not consider the $K\Sigma$
photoproduction data to avoid the need to also determine the
parameters associated with the photo-excitations of  about 12
other isospin $I=3/2$ $N^*$ states. Obviously, our results
 serve as a good starting point for a subsequent investigation
including all $KY$ channels. Our results in that direction will be
published elsewhere.

This paper is organized as follows: In Section II, the theoretical
frame is presented. The main content of our coupled-channel
formalism is then given, followed by an outline of the relevant
constituent quark model for the direct $\gamma p~\to~K^+ \Lambda$
channel. There, the novelties of our approach are discussed.
Section III is devoted to numerical results and comparisons with
available data for $\pi^- p \to K^\circ \Lambda,~K^\circ
\Sigma^\circ$ and $\gamma p \to K^+ \Lambda$. For this latter
reaction, the most relevant known nucleon resonances are singled
out and possible manifestations of new baryon resonances are
discussed. Meson cloud effects are exhibited by examining
multipoles from the obtained model. Summary and conclusions are
reported in Section IV.


\section{Theoretical Formulation}
\label{sec:tf}
In this Section, we first present our dynamical coupled-channel
approach for the photoproduction process including intermediate
$\pi N$ and $KY$ channels. Then, we outline the constituent quark
model used for the direct $KY$ photoproduction reaction.

\subsection{Coupled-channel formalism}
\label{subsec:cc} The coupled-channel approach presented here is
derived from a general formulation reported in
Refs.~\cite{lms,bl}. The starting point is a Hamiltonian
consisting of non-resonant terms $v_{a,b}$ plus resonant terms
$v^R_{a,b}={\Gamma^{\dagger}_{N^*,a}
\Gamma_{N^*,b}}/{(E-M^0_{N^*})}$,
where $a,b$ are the considered meson-baryon channels, $M^0_{N^*}$
is the bare mass of the $N^*$ state, and $\Gamma_{N^*,a}$ describe
the $N^*\to a$ decays. Such a Hamiltonian can be derived from
effective Lagrangians using a unitary transformation method
developed in Ref.~\cite{SatoLee}.
By using the two potential formulation~\cite{gold}, as also derived
explicitly in Appendix A,  one can cast exactly 
the transition amplitude $T_{a,b}(E)$ for the $a\to b$ reaction
into a sum of non-resonant $t_{a,b}(E)$ and resonant
$t^{R}_{a,b}(E)$ terms:
\begin{eqnarray}
T_{a,b}(E) &=& t_{a,b}(E) + t^{R}_{a,b}(E) \, . \label{t1}
\end{eqnarray}
The first term of Eq.~(\ref{t1}) is determined only by the
non-resonant interactions
\begin{eqnarray}
t_{a,b}(E)= v_{a,b} +\sum_{c} v_{a,c}\ G_c(E)\  t_{c,b}(E)\,,
\label{t3}
\end{eqnarray}
where $G_c(E)$ is the propagator of the meson-baryon state $c$.
The resonant term is
\begin{eqnarray}
t^{R}_{a,b}(E) &=& \sum_{N^*_i, N^*_j}
\bar{\Gamma}^\dagger_{N^*_i, a}(E)\  [G^{N*}(E)]_{i,j} \
\bar{\Gamma}_{N^*_j, b}(E) \,. \label{t2}
\end{eqnarray}
The resonant amplitude in Eq.~(\ref{t2}) is determined by the
dressed vertex
\begin{eqnarray}
\bar{\Gamma}_{N^*,a}(E) &=&
  { \Gamma_{N^*,a}} + \sum_{b} \Gamma_{N^*,b}\
G_{b}(E)\ t_{b,a}(E)\,, \label{t4}
\end{eqnarray}
and the dressed $N^*$ propagator
\begin{eqnarray}
[G^{N^*}(E)^{-1}]_{i,j}(E) = (E - M^0_{N^*_i})\delta_{i,j} -
\Sigma_{i,j}(E) \,. \label{t5}
\end{eqnarray}
Here the $N^*$ self-energy is defined by
\begin{eqnarray}
\Sigma_{i,j}(E)= \sum_{a}\bar{\Gamma}_{N^*_i,a} G_{a}(E)
\Gamma^\dagger_{N^*_j,a}(E)\,. \label{t6}
\end{eqnarray}

In this work, we make the following simplifications. We keep only
three channels $\gamma N$, $KY$ and $\pi N$, and
 neglect the terms with 
electromagnetic coupling strenghts higher than the first order $e$.
 We further assume
that the $N^*$ propagator Eq.~(\ref{t5}) can be replaced by a
simple phenomenological Breit-Wigner form.
Then, Eqs.~(\ref{t1})-(\ref{t6}) are reduced to the following
expressions for calculating the $\gamma N \to KY$ and $\pi N \to
KY$ amplitudes:
\begin{eqnarray}
T_{\gamma N, KY}(E) &=& t_{\gamma N,KY}(E) + \sum_{N^*}
 \frac{\bar{\Gamma}^{\dagger}_{N^*,\gamma N}\bar{\Gamma}_{N^*,KY}}
{E- M_{N^*} +i{\Gamma^{tot}(E)}/{2}} \,,
\label{gk1a} \\
T_{\pi N, KY}(E) &=& t_{\pi N,KY}(E) + \sum_{N^*}
 \frac{\bar{\Gamma}^{\dagger}_{N^*,\pi N}\bar{\Gamma}_{N^*,KY}}
{E- M_{N^*} +i{\Gamma^{tot}(E)}/{2}},
\label{gk1b}
\end{eqnarray}
with
\begin{eqnarray}
\bar{\Gamma}^{\dagger}_{N^*,\gamma N}
&=&{\Gamma}^{\dagger}_{N^*,\gamma N} +[ t_{\gamma N, KY}G_{KY}
{\Gamma}^{\dagger}_{N^*,KY} +t_{\gamma N, \pi N}G_{\pi N}
{\Gamma}^{\dagger}_{N^*,\pi N}]\,, \label{gk2a}\\
\bar{\Gamma}^{\dagger}_{N^*,\pi N}
&=&{\Gamma}^{\dagger}_{N^*,\pi N} +[ t_{\pi N, KY}G_{KY}
{\Gamma}^{\dagger}_{N^*,KY} +t_{\pi N, \pi N}G_{\pi N}
{\Gamma}^{\dagger}_{N^*,\pi N}] \,,
\label{gk2b} \\
\bar{\Gamma}_{N^*,KY}
&=& {\Gamma}_{N^*,KY}+[ {\Gamma}_{N^*,KY}G_{KY} t_{KY,KY}
+{\Gamma}_{N^*,\pi N}G_{\pi N}t_{\pi N, KY}] \,.
\label{gk2c}
\end{eqnarray}

It is clear that the first step to solve the above equations is to
develop models for calculating all non-resonant amplitudes. To
first order in electromagnetic coupling,  within the considered
$\gamma N \oplus KY \oplus \pi N$ space, Eq.~(\ref{t3}) leads to
\begin{eqnarray}
t_{\gamma N, KY} &=& v_{\gamma N, KY} [ 1 + G_{KY}(E)t_{KY,KY}(E)]
+v_{\gamma N, \pi N} G_{\pi N}(E) t_{\pi N, KY} \nonumber  \\
&=& v_{\gamma N, KY} [ 1 + G_{KY}(E)t_{KY,KY}(E)]
+t_{\gamma N, \pi N} G_{\pi N}(E) v_{\pi N, KY}.
\label{gk3}
\end{eqnarray}
Here we note that the second line of the above equation is
obtained from using the well-known property $vgt = tgv$. The
non-resonant amplitudes $t_{KY,KY}$ and $t_{\pi N, KY}$ in
Eq.~(\ref{gk3}) are obtained by solving Eq.~(\ref{t3}) within the
subspace $KY \oplus \pi N$. For numerical reasons, we follow the
procedure of Ref.~\cite{CC-01} to eliminate $t_{\pi N,\pi N}$ from
these coupled equations. We then obtain the following equations
\beqa
t_{KY, KY} &=& v^{\rm eff}_{KY, KY}
            + \sum_{KY} \;  v^{\rm eff}_{KY, KY}\ G_{KY}\ t_{KY, KY},
\label{gk4a} \\
t_{KY, \pi N} &=& [v_{KY, \pi N}
              + t_{KY, KY}\ G_{KY }\ \;v_{KY, \pi N}] \nonumber \\
              && \times~ [1 + G_{\pi N}\ \hat{t}_{\pi N,\pi N}],
\label{gk4b}
\eeqa
where
\beqa v^{\rm eff}_{KY, KY}   &=& v_{ KY, KY}+ \sum_{\pi N} \;
v_{KY, \pi N}\ G_{\pi N}\ v^{\rm eff}_{\pi N, KY}, \label{gk5}
\eeqa
with
\beqa
v^{\rm eff}_{\pi N, KY}&=& v_{\pi N, KY}
            + \sum_{\pi N}\; \hat{t}_{\pi N, \pi N}\ G_{\pi N}\ \;v_{\pi N, KY}.
\label{gk6}
\eeqa
The pure $\pi N$ scattering t-matrix $\hat{t}_{\pi N, \pi N}$ in the
above equations is defined by
\beqa \hat{t}_{\pi N, \pi N}=v_{\pi N, \pi N}+v_{\pi N, \pi N}\
G_{\pi N}\ \hat{t}_{\pi N, \pi N} \,. \label{gk7} \eeqa
We see that Eqs.~(\ref{gk4a}) and (\ref{gk7}) are single channel
integral equations. The couplings between $\pi N$ and $KY$
channels are isolated in the effective potentials
$v^{eff}_{KY,KY}$ and $v^{eff}_{\pi N, KY}$. Clearly, the use of
Eqs.~(\ref{gk4a})-(\ref{gk7}) greatly simplifies the numerical
task of handling the matrix problem associated with the original
coupled-channel integral equations in the subspace $KY \oplus \pi
N$. In fact, this technique will be useful for future
investigations including additional channels.

To solve the above equations, we employ the non-resonant
potentials $v_{KY,KY}$, $v_{\pi N, KY}$ derived in
Ref.~\cite{CC-04} from effective Lagrangians using a unitary
transformation method of Ref.~\cite{SatoLee}. The expressions for
these potentials can be found there and will not be repeated here.
However, we depart from Ref.~\cite{CC-04} in two aspects. First,
Eq.~(17) for determining $\hat{t}_{\pi N,\pi N}$ was not solved
directly in Ref.~\cite{CC-04}. Instead, it was estimated from
using the empirical $\pi N \to \pi N$ amplitudes.
In this work, we solve Eq.~(\ref{gk7}) by using $v_{\pi N,\pi N}$ of
Ref.~\cite{SatoLee} which was also derived from effective Lagrangians
using the same unitary transformation method.
The second new aspect of our calculations is to include the distortion
effects on the $N^*$ decays, defined by the term within the square brackets
in the right-hand-side of Eqs.~(\ref{gk2b})-(\ref{gk2c}), which were
neglected in the calculations of Ref.~\cite{CC-04}.
It turns out that these two refinements do not change much the quality
of the fits to the $\pi N \to KY$ data.
More details will be given in the next Section.

We now discuss the calculation of the non-resonant kaon
photoproduction amplitude defined by Eq.~(\ref{gk3}). While the main
contribution to $t_{\gamma N, KY}$ is expected to be from the
direct transition amplitude $v_{\gamma N, KY}$, the calculations
of the coupled-channel effects due to the $\pi N$ channel require
a model for the $\gamma N \to \pi N$ amplitude $t_{\gamma N, \pi
N}$. The amplitude $t_{\gamma N, \pi N}$ is expected to be rather
complicated in the second and third $N^*$ regions. Full
construction of $t_{\gamma N, \pi N}$ is far beyond the scope of
this work. To make progress, we follow the phenomenological
procedure of Ref.~\cite{RPF} to define $t_{\gamma N,\pi N}$ in
terms of the empirical $\gamma N \to \pi N$ amplitude and the
resonant amplitude constructed from the quark model predictions of
Ref.~\cite{Cap92,Cap94,Cap98}. Explicitly, we define
\beq
t_{\pi N, \gamma N}\equiv T^{\rm SAID}_{\pi N, \gamma N}
                             - t^{\rm QM,R}_{\pi N, \gamma N} ,
\label{eq:pionphoto}
\eeq
where $t^{\rm QM,R}_{\pi N, \gamma N}$ is the quark model
amplitude given explicitly in Ref.~\cite{RPF} and $T^{\rm
SAID}_{\pi N, \gamma N}$ is obtained from the 1995 solution of the
SAID~\cite{SAID} analysis. As an alternative, we can replace
$t^{\rm QM,R}_{\pi N, \gamma N}$ by $t^{\rm PDG,R}_{\pi N, \gamma
N}$ which is the $\gamma N \to N^* \to \pi N$ amplitude defined by
the resonance parameters listed by PDG. Unfortunately, the
parameters of $\gamma N \to N^*$ for most of the considered $N^*$
are not well determined by PDG. In fact, this work is one of the
possible ways to learn about these $\gamma N \to N^*$ amplitudes
by considering the photoproduction channels other than the $\pi N$
channel. We thus use Eq.~(\ref{eq:pionphoto}) in this work.

Eq.~(\ref{eq:pionphoto}) only defines the on-shell values of the
amplitude $t_{\pi N, \gamma N}$. For the calculation of
Eq.~(\ref{gk3}), which involves integrations over off-shell matrix
elements, we define the following off-shell behavior
\begin{eqnarray}
t_{\pi N, \gamma N}(q,k_0,W) = t_{\pi N, \gamma N}(q_0,k_0,W)
\frac{F(q,\Lambda)}{ F(q_0,\Lambda)}\,,
\end{eqnarray}
with
\begin{eqnarray}
F(q,\Lambda)&=& \left({\Lambda^2\over \Lambda^2+q^2}\right)^2\,,  \\
q_0&=&{[{(W^2-m_N^2-m_\pi^2)^2 -4 m_N^2 m_\pi^2}]^{1/2}\over 2 W}\,,
\end{eqnarray}
where $W$ is the invariant mass of the $\pi N$ system, $q$ is $\pi
N$ off-shell momentum, $k_0$ is the on-shell momentum of the
initial $\gamma N$ system, and the cutoff $ \Lambda$ is an
adjustable parameter in our fit to the $\gamma N \to KY$ data. We
find $\Lambda$ = 1.5 GeV.
%


\subsection{Direct channel}
\label{subsec:dc}

For the non-resonant $\gamma N \to KY$ transition amplitude
$v_{\gamma N, KY}$ and the resonant amplitude, we follow the
procedure of Refs.~\cite{CQM-1a,CQM-1b}. The details can be found
there and will not be repeated here. Below, we summarize the main
points needed in the subsequent Section.

The chiral constituent quark approach is based on a low energy
QCD-inspired Lagrangian~\cite{QCD}, where the 
scattering matrix for the photoproduction of pseudoscalar
mesons can be derived~\cite{Li97} as
\begin{eqnarray}\label{eq:Mfi}
{\mathcal M}_{fi}&=&\langle N_f| H_{m,e}|N_i \rangle +
\sum_j\bigg \{ \frac {\langle N_f|H_m |N_j\rangle
\langle N_j |H_{e}|N_i\rangle }{E_i+\omega-E_j}+
\frac {\langle N_f|H_{e}|N_j\rangle \langle N_j|H_m
|N_i\rangle }{E_i-\omega_m-E_j}\bigg \}  \nn \\
&& + {\mathcal M}_T.
\end{eqnarray}
Here, $N_i(N_f)$ is the initial (final) state of the nucleon,
$\omega (\omega_{m})$ represents the energy of incoming (outgoing)
photons, and $H_m$ and $H_e$ are pseudovector and electromagnetic
couplings, respectively, and $N_j$ is the intermediate baryon.

The first term in Eq.~(\ref{eq:Mfi}) is a seagull term. The second
and third terms correspond to the {\it s-} and {\it u-}channels,
respectively. The last term  ${\mathcal M}_T$ is the {\it
t-}channel contribution.

The contribution from  the {\it s-}channel resonances to the transition
matrix elements can be written as
\begin{eqnarray}\label{eq:MR}
{\mathcal M}^{CQM}_{N^*}=\frac
{2M_{N^*}}{W^2-M_{N^*}(M_{N^*}-i\Gamma(q))}\  e^{-\frac
{{k}^2+{q}^2}{6\alpha^2_{ho}}}\ {\mathcal A}_{N^*},
\end{eqnarray}
with  $k=|\vk|$ and $q=|\vq|$ the momenta of the incoming photon
and the outgoing meson, respectively; W is the total energy of the
system, $e^{- {({k}^2+{q}^2)}/{6\alpha^2_{ho}}}$ a form factor in
the harmonic oscillator basis with the parameter $\alpha^2_{ho}$
related to the harmonic oscillator strength in the wave-function,
and $M_{N^*}$ and $\Gamma(q)$ the mass and the total width of the
resonance, respectively.  The amplitudes ${\mathcal A}_{N^*}$ are
divided into two parts: the contribution from each resonance below
2 GeV (these transition amplitudes  have been translated into the
standard CGLN amplitudes in the harmonic oscillator basis), and
the contributions from the resonances above 2 GeV , which are
treated as degenerate~\cite{Li97}.

The contributions from each resonance is determined by
introducing~\cite{CQM-1a} a new set of real parameters $C_{{N^*}}$
for the amplitudes ${\mathcal A}_{{N^*}}$:
\begin{eqnarray}\label{eq:AR}
{\mathcal A}_{N^*} \to C_{N^*} {\mathcal A}_{N^*} ,
\end{eqnarray}
so that
\begin{eqnarray}\label{MRexp}{\mathcal M}_{N^*}^{exp} = C^2_{N^*}
 {\mathcal M}_{N^*}^{CQM} ,
\end{eqnarray}
where ${\mathcal M}_{N^*}^{exp}$ is the experimental value of the
observable, and ${\mathcal M}_{N^*}^{CQM}$ is calculated in the
quark model~\cite{CQM-1b}. For instance, for resonance with mass
$\leq$~2 GeV, the $SU(6)\otimes O(3)$ symmetry predicts
$C_{N^*}$~=~0.0 for ${S_{11}(1650)} $, ${D_{13}(1700)}$, and
${D_{15}(1675)} $ resonances, and $C_{N^*}$~=~1.0 for other ones.
However, deviations from those central values are anticipated
within the broken $SU(6)\otimes O(3)$ symmetry, due for example to
one-gluon exchange mechanisms~\cite{Mixing}.

%
\section{Results and Discussion}
\label{sec:rd}
This Section is devoted to the application of our formalism to the
production of kaon-hyperon final states in $\pi N$ and $\gamma p$
collisions.

To that end, we need first to study $\pi N \rightarrow  KY$, and
$KY \to KY$ processes. In the following we first compare our $\pi
N~\to~KY$ results with the relevant data and also extract $N^*$
information within the considered model. Then we present results
for the photoproduction channel and discuss issues related to the
missing resonances.

\subsection{ $\pi N \to KY$ Reaction}
 \label{sec:pinky}
As seen in Eq.~(\ref{gk3}),  to calculate $\gamma N \to KY$
amplitude our first step is to construct the non-resonant
amplitudes $t_{KY,KY}$ and $t_{\pi N, KY}$. These are obtained
within our model by solving the coupled-channel equations
(\ref{gk4a})-(\ref{gk7}).
The input of these coupled-channel equations are the potentials
$v_{KY,\pi N}$, $v_{KY,KY}$, and an effective non-resonant amplitude
$\hat{t}_{\pi N\pi N}$ which is defined by Eq.~(\ref{gk7}).
The parameters of these potentials are then adjusted along with the
$N^*$ parameters associated with the resonant term of Eq.~(\ref{gk1b})
to fit the $\pi^- p \to K^\circ \Lambda$ and
$\pi^- p \to K^\circ \Sigma^\circ$
data~\cite{Bak78L,Kna75,Sax80,Bak78S,Har80}.


{\squeezetable
\begin{table}[tb]
\centering
\begin{tabular}{cccr}
\hline
\hline
  \ Notation\ &\ Resonance\   &\  Coupling\         &\ Value  \        \\
\hline

       &                      & $f_{K\Lambda N}$       &  -0.61        \\
       &                      & $f_{K\Sigma N}$        &   0.12       \\
       &                      & $f_{\pi\Sigma\Lambda}$ &   0.08        \\
       &                      & $f_{\pi\Sigma\Sigma}$  &   0.00        \\
  $N4$ & $S_{11}(1650)~1/2^-$ & $f_{K\Lambda N4}$      &  -0.25       \\
       &                      & $f_{K\Sigma N4}$       &  -0.20       \\
  $N5$ & $D_{13}(1700)~3/2^-$ & $f_{K\Lambda N5}$      &  -0.33        \\
       &                      & $f_{K\Sigma N5}$       &   0.08        \\
  $N6$ & $P_{11}(1710)~1/2^+$ & $f_{K \Lambda N6}$     &   0.09        \\
       &                      & $f_{K\Sigma N6}$       &  -0.32        \\
  $N7$ & $P_{13}(1720)~3/2^+$ & $f_{K\Lambda N7}$      &  -0.56       \\
       &                      & $f_{K\Sigma N7}$       &   0.54       \\[1ex]
  $D1$ & $S_{31}(1900)~1/2^-$ & $f_{K\Sigma D1}$       &   0.09         \\
  $D2$ & $P_{31}(1910)~1/2^+$ & $f_{K\Sigma D2}$       &   0.20        \\
  $D3$ & $P_{33}(1920)~3/2^+$ & $f_{K\Sigma D3}$       &  -0.20        \\
  $L3$ & $S_{01}(1670)~1/2^-$ & $f_{\pi\Sigma L3}$     &  -0.20        \\
  $L5$ & $P_{01}(1810)~1/2^+$ & $f_{\pi\Sigma L5}$     &  -0.01      \\
  $S1$ & $P_{11}(1660)~1/2^+$ & $f_{\pi\Lambda S1}$    &  -0.20       \\
       &                      & $f_{\pi\Sigma S1}$     &  -0.20       \\
  $S4$ & $D_{13}(1670)~3/2^-$ & $f_{\pi\Lambda S4}$    &   0.22       \\
       &                      & $f_{\pi\Sigma S4}$     &   0.05       \\
\hline

       &  $K^*NY$ couplings   &  $f_{K^*N\Lambda}^V$  &    0.71      \\
       &                      &  $f_{K^*N\Lambda}^T$  &   -3.97      \\
       &                      &  $f_{K^*N\Sigma}^V$   &   -0.53      \\
       &                      &  $f_{K^*N\Sigma}^T$   &    0.52      \\
\hline
       &   cut-offs           &  $\Lambda_s$          &   623.0      \\
       &                      &  $\Lambda_u$          &  1468.0      \\
       &                      &  $\Lambda_i$          &   930.0      \\
       &                      &  $\Lambda_{\pi N}$    &  1491.0      \\
\hline
       &  off-shell           &  $X$                  &     2.0\\
\hline
       &  Reduced $\chi^2$    &                       &     1.86\\
\hline
\hline
\end{tabular}
\caption {Coupling constants in $\pi N \to KY$
and $KY \to KY$. The values are extracted from
our minimization procedure. The parameters are defined in
the model B of Ref.~\cite{CC-04}.}
\protect\label{tab:mbparam}
\end{table}
}

This policy was pursued in Ref.~\cite{CC-04} but with the
simplifications that the distortion factors, the terms within the
square brackets in Eqs.~(\ref{gk2b})-(\ref{gk2c}), were not
included in calculating the resonant term of Eq.~(\ref{gk1b}).
Furthermore, the non-resonant $\hat{t}_{\pi N,\pi N}$ defined by
Eq.~(\ref{gk7}) was only roughly estimated using the empirical
$\pi N$ amplitude.

\begin{figure}[th]
\vspace{25pt}
\begin{center}
\mbox{\epsfig{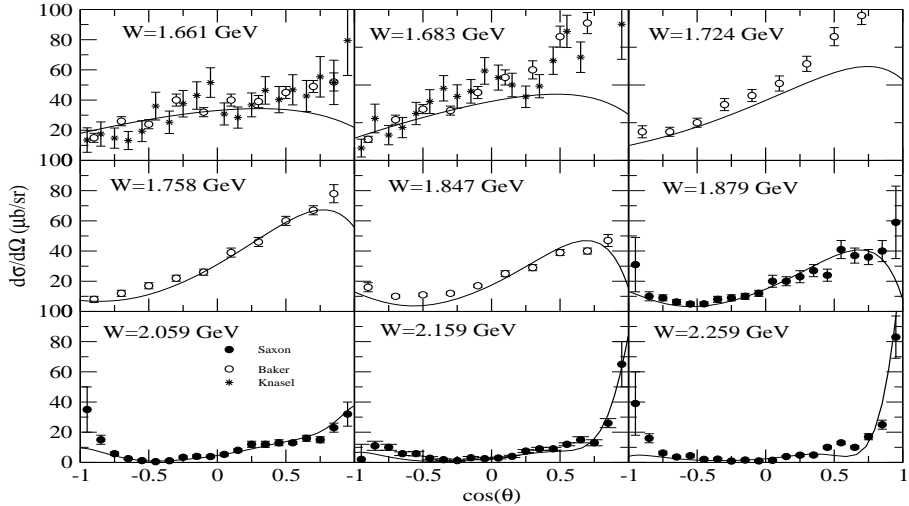}}
\end{center}
\caption{Differential cross-section for the reaction
$\pi^- p \to K^\circ \Lambda$. The solid curves are from the fits using the
coupled-channel model of this work.
Data are from Refs.~\cite{Bak78L,Kna75}.}
\protect\label{fig:dsLa}
\end{figure}
\begin{figure}[hbt]
\vspace{25pt}
\begin{center}
\mbox{\epsfig{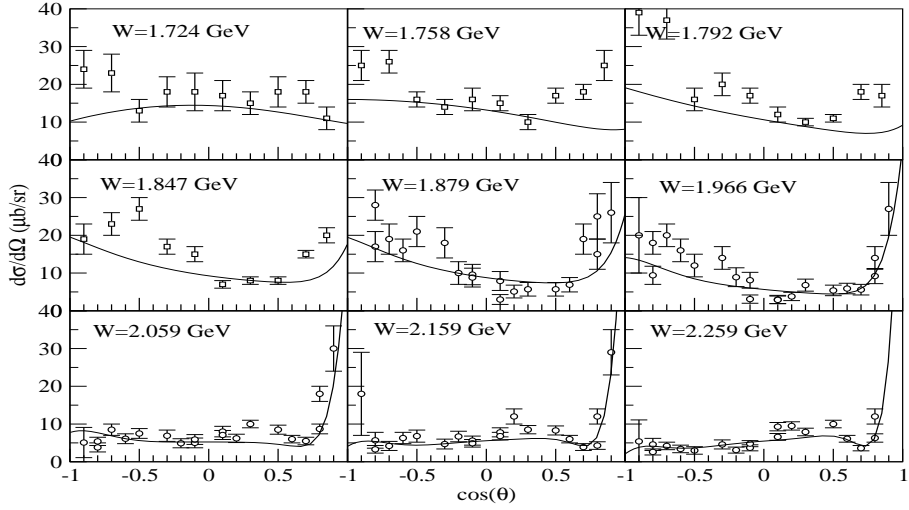}}
\end{center}
\caption{Differential cross-section for the reaction
$\pi^- p \to K^\circ \Sigma^\circ$.
The solid curves are from the fits using the coupled-channel
model of this work.
Data are from
Refs.~\cite{Bak78S,Har80}.}
\protect\label{fig:dsS}
\end{figure}

In this work, we have corrected these two deficiencies as discussed in
Section II.A and thus have refined the potentials $v_{\pi N, KY}$ and
$v_{KY,KY}$ and the relevant $N^*$ parameters.

The fitting procedure is explained in detail in Section III of
Ref.~\cite{CC-04}. Here we recall a few points to make the present
Section self-consistent. In that paper, we classified the
parameters in three sets (Tables I to III in Ref.~\cite{CC-04}).
Set I includes 9 couplings, the values of which are taken from the
SU(3)-symmetry predictions or PDG partial decay widths; namely,
$f_{\pi NN}$,
$f_{\pi N N^*}$, $f_{\pi N \Lambda^*}$, and $f_{\pi N \Sigma^*}$, with
$N^*~\equiv~S_{11}(1650),~D_{13}(1700),~P_{11}(1710),~P_{13}(1720)$,
$\Lambda^*~\equiv~S_{01}(1670),~P_{01}(1810)$, and
$\Sigma^*~\equiv~P_{11}(1660),~D_{13}(1670)$.

The adjustable parameters are in the remaining two sets. Set II
 includes the following coupling constants: $f_{KYN}$,
$f_{KY N^*}$, $f_{KY \Delta ^*}$, $f_{\pi YY}$, and $f_{\pi
YY^*}$. The values extracted for those parameters within the
present work are given in Table~\ref{tab:mbparam}, rows 2 to 22.
Here we follow  model B of Ref.~\cite{CC-04} by allowing the
parameters of the model to vary by $\pm 20 \%$ around the central
values taken from PDG~\cite{PDG} and/or from quark
model~\cite{Cap92,Cap94} predictions. Finally rows 23 to 31 in
Table~\ref{tab:mbparam} correspond to  Set III in
Ref.~\cite{CC-04}.

In Figs.~\ref{fig:dsLa} to \ref{fig:asyS}, the results of our
model are compared with the differential cross-section and recoil
hyperon polarization data~\cite{Bak78L,Bak78S,Kna75,Sax80,Har80}
for processes $\pi^- p~\to~K^\circ \Lambda$ and $\pi^-
p~\to~K^\circ \Sigma^\circ$.

In Figs.~\ref{fig:dsLa} and~\ref{fig:dsS} we show the quality of
our fits to the differential cross section data for $\pi^- p \to
K^\circ \Lambda$ and $\pi^- p \to K^\circ \Sigma^\circ$,
respectively. In Figs.~\ref{fig:asyLa} and~\ref{fig:asyS} our
results for the asymmetry data for the same reactions are
depicted. The acceptable agreement between model and data, as well
as $\chi^2_{d.o.f}$, compare well with our previous
results~\cite{CC-04}. Nevertheless, we consider the present model
slightly more reliable than the model B in Ref.~\cite{CC-04}.
Actually, some of the coupling constants, Table~\ref{tab:mbparam},
get (much) closer to constituent quark model values~\cite{Cap98},
e.g. $f_{K \Lambda D_{13}(1700)}$, $f_{K \Sigma D_{13}(1700)}$,
$f_{K \Lambda P_{11}(1710)}$, $f_{K \Lambda P_{13}(1720)}$, $f_{K
\Sigma S_{31}(1900)}$, and $f_{K \Sigma P_{33}(1920)}$.

\begin{figure}[t]
\vspace{25pt}
\begin{center}
\mbox{\epsfig{file=fig3, width=120mm, height=65mm}}
\end{center}
\caption{$\Lambda$ recoil polarization asymmetries for the
reaction $\pi^- p \to K^\circ {\vec \Lambda}$.
The solid curves are from the fits using the coupled-channel
model of this work.
 Data are
from Refs.~\cite{Bak78L,Sax80}.}
\protect\label{fig:asyLa}
\end{figure}

\begin{figure}[ht]
\begin{center}
\mbox{\epsfig{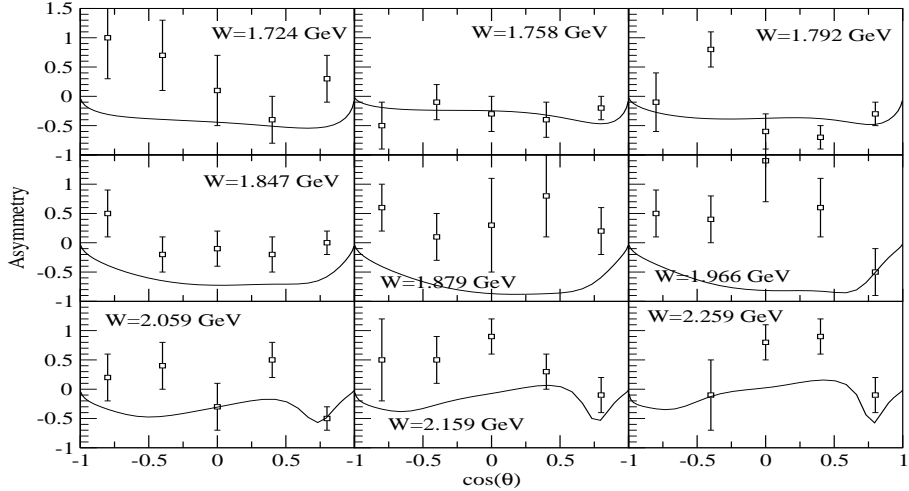}}
\end{center}
\caption{$\Sigma$ recoil polarization asymmetries for the
reaction $\pi^- p \to K^\circ {\vec \Sigma^\circ}$.
The solid curves are from the fits using the coupled-channel
model of this work.
Data are
from Ref.~\cite{Har80}.}
\protect\label{fig:asyS}
\end{figure}

In summary, it turns out that the aforementioned improvements do
not change much with respect to our previous model~\cite{CC-04}.
For the $KY~\to~KY$ processes, we also get comparable results to
those reported in the latter paper. There is no data for $KY \to
KY$ scattering to test our model. 
The situation might change in the near future with the
advent of highly
accurate data from the EPECUR~\cite{epecur} and
J-PARC~\cite{JPARC} projects. Those data
 will certainly afford deeper insights
into the meson-baryon interactions. However, the results shown in
Figs.~\ref{fig:dsLa}-\ref{fig:asyS} are sufficient for the purpose
of studying coupled-channel effects.


\subsection{ $\gamma p \to K^+ \Lambda$ Reaction}
\label{ssec:klp}

We have performed a thorough study of all the latest relevant data
(Table~\ref{tab:obs-1}). The data released in December 2005 by
LEPS~\cite{LEPS05} are not included in our fitted data base.
However, they are depicted in the relevant figures below and
briefly discussed.

The strong interaction channels amplitudes $v_{\pi N, KY}$ and
$t_{KY,KY}$ are determined above, and $t_{\gamma N, \pi N}$
computed  from Eq.~(\ref{eq:pionphoto}).

\begin{table}[ht]
\begin{tabular}{llrrrr}
\hline
\hline
  Lab/Collaboration & Observable && \# of data points && Ref. \\
\hline
ELSA/SAPHIR      & Differential cross-section    &&  720 && \cite{ELSA}\\
JLab/CLAS        & Differential cross-section    && 1068 && \cite{JLab05} \\
JLab/CLAS        & Recoil polarization asymmetry &&  233 && \cite{JLab04} \\
SPring-8/LEPS    & Polarized beam asymmetry      &&   44 && \cite{LEPS}\\
Bonn synchrotron & Polarized target asymmetry    &&    3 && \cite{Target}\\
\hline
\hline
\end{tabular}
\caption{Data sets investigated in the present work. Here, we have not included
268 cross-section data points from CLAS \cite{JLab05} for $E_{\gamma} >$ 2.6 GeV
(W~$\ge$~2.4 GeV),
in order to concentrate on the baryon resonances energy range.}
\protect\label{tab:obs-1}
\end{table}

For both non-resonant and resonant $\gamma p \to K^+ \Lambda$ amplitudes
we use a constituent quark model~\cite{CQM-1b,BS04}.
We recall that the resonant term of Eq.~(\ref{gk1a}) contains a term
\begin{eqnarray}
t^R_{\gamma N , KY} = v^R_{\gamma N,KY}[1+G_{KY}\ t_{KY,KY}],
\end{eqnarray}
with
\begin{eqnarray}
v^R_{\gamma N,KY}=\frac{\Gamma^{\dagger}_{N^*,\gamma N}\
\Gamma_{N^*,KY}} {E- M_{N^*} +i{\Gamma^{tot}(E)}/{2}} .
\end{eqnarray}
To use the $N^*$ contributions defined by
Eqs.~(\ref{eq:MR})-~(\ref{MRexp}), we replace the above
expression by
\begin{eqnarray}
v^R_{\gamma N, KY} = C_{N^*}M^{CQM}_{N^*}, \label{eq:saghaili}
\end{eqnarray}
where $M^{CQM}_{N^*}$ is calculated~\cite{CQM-1b} from the
constituent quark model. The $SU(6)\otimes O(3)$ symmetry breaking
coefficient $C_{N^*}$, Eq.~(\ref{eq:saghaili}), are treated as
constrained adjustable parameters~\cite{CQM-1a,CQM-2} in fitting
the data.

%

\subsubsection{Model search}
 \label{sec:param}

In this Section,  we explain the procedure used to build a model
for all available data. 

Here we would like to emphasize that the
CLAS~\cite{JLab04} and SAPHIR~\cite{ELSA} data released in 2004,
with some 2000 data points for differential cross-sections, showed
significant discrepancies with each other. This fact led the
phenomenologists either to concentrate on one of the two sets or
produce one model per data set. This uncomfortable situation is
now significantly cured thanks to the CLAS Collaboration's new
data~\cite{JLab05}, made available in 2005.
In an earlier attempt~\cite{CC-05b}, we underlined this improvement
in experimental data base and reported our preliminary results
obtained with respect to both SAPHIR 2004 and CLAS 2005 data.

Table~\ref{tab:obs-1} summarizes the
content of the data base used to determine the adjustable
parameters of our approach; namely, known resonances strengths. 
Additional parameters due to the introduction of new resonances 
are discussed in Section~\ref{sec:new-res}.
Differential cross-section data provide, of course, the main
constraints on the model ingredients. Consequently, our starting
point was to fit separately the CLAS and SAPHIR cross-section
data, for which the reduced $\chi^2$s turned out to be 2.1 and
1.3, respectively. The significantly larger $\chi^2_{d.o.f}$ found
using the CLAS data is due to their smaller uncertainties compared
to those of SAPHIR data. However, this fact might not be the only
source of the difference in  $\chi^2$s.  Actually, two
considerations are in order here:

i) The earlier data from CLAS~\cite{JLab04} showed significant
discrepancies with SAPHIR~\cite{ ELSA} data. Although the new
CLAS~\cite{JLab05} data set has significantly reduced those
discrepancies, in some phase space regions results from the two
data set differ still by more than 2$\sigma$;

ii) The strengths of
resonances, which constitute our main adjustable parameters, are
rather tightly constrained by $SU(6)\otimes O(3)$ symmetry.
Consequently, the fact that we obtain a much better
$\chi^2_{d.o.f}$ for the SAPHIR data compared to the one for the
CLAS data leads to the conclusion that our approach is more in
line with the SAPHIR differential cross-section data than with
CLAS results.

Keeping the above considerations in mind, we present two models
here:

\noindent i) Model $M_1$: all SAPHIR and most recent CLAS
differential cross-sections (first two rows in Table II) were
fitted simultaneously.

\noindent ii) Model $M_2$: all cross-section and polarization asymmetries
(Table~\ref{tab:obs-1}) were fitted simultaneously.

Extracted values for the eleven adjustable parameters are given in
Table~\ref{tab:cqm-1}. That Table contains the $KYN$ coupling
constant and the strengths of known resonances with masses $\le$ 2
GeV. The higher-mass, known resonances are treated as degenerate
in a compact way~\cite{CQM-1b,Li97} and bear no symmetry breaking
coefficients. Moreover, the Roper resonance, although explicitly
present in our approach, does not contribute to the reaction
mechanism due to its low mass with respect to the reaction
threshold. In addition to those known resonances, we also
introduce 3 adjustable parameters per each of  three newly
proposed $S_{11}$, $P_{13}$, and $D_{13}$ resonances, as discussed
below (see Table~\ref{tab:cqm-2} in Section~\ref{sec:new-res}).

\begin{table}[htb]
\begin{tabular}{lrrrr}
\hline
\hline
  Parameter       && Model $M_1$ && Model $M_2$ \\
\hline
$g_{KN \Lambda}$  &&    8.02     &&  8.00  \\
$C_{S_{11}(1535)}$&&   -0.85     && -0.82  \\
$C_{S_{11}(1650)}$&&   -0.10     && -0.22  \\
$C_{P_{11}(1710)}$&&    1.79     && -1.08  \\
$C_{D_{13}(1520)}$&&   -2.00     && -2.00  \\
$C_{D_{13}(1700)}$&&    0.16     && -0.19  \\
$C_{P_{13}(1720)}$&&   -0.40     &&  0.05  \\
$C_{P_{13}(1900)}$&&    0.80     &&  1.60  \\
$C_{D_{15}(1675)}$&&   -0.09     &&  0.22  \\
$C_{F_{15}(1680)}$&&    1.43     &&  1.99  \\
$C_{F_{15}(2000)}$&&    1.28     &&  1.59  \\
\hline
$\chi^2_{d.o.f}$  &&    2.49     &&  3.32  \\
\hline
\hline
\end{tabular}
\caption{Kaon-nucleon-hyperon coupling constant, $SU(6)\otimes O(3)$ symmetry
breaking coefficient $C_{N^*}$ as in Eq.~(\ref{eq:saghaili}), and reduced $\chi^2$
for models $M_1$ and $M_2$.}
\protect\label{tab:cqm-1}
\end{table}


\begin{figure}[htb]
\vspace{30pt}
\begin{center}
\mbox{\epsfig{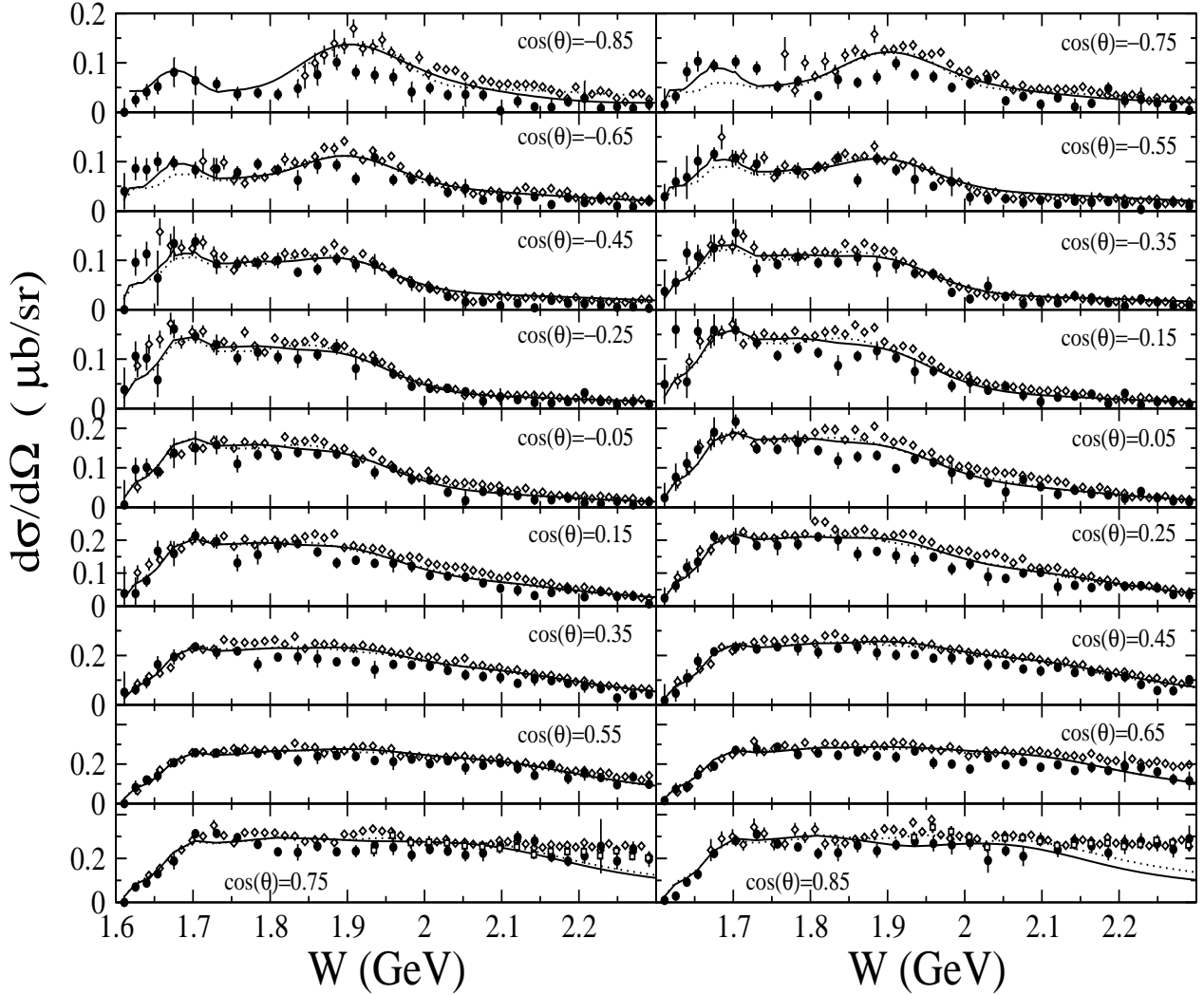}}
\end{center}\caption{Differential cross-section for the reaction $\gamma p \to
K^+ \Lambda$ as a function of total center-of-mass energy.
Dotted and solid curves correspond to models $M_1$ and
$M_2$ respectively. Data are from Ref.~\cite{JLab05} (open
diamonds), Ref.~\cite{ELSA} (full circles), and LEPS~\cite{LEPS05}
(open squares in the cells corresponding to cos$\theta$ = 0.75 and
0.85). Plotted data from Ref.~\cite{JLab05} are measured at 
$cos(\theta)'=cos(\theta)+0.05$.}
\protect\label{fig:dxs-1}
\end{figure}

Here we would like to comment about the extracted values of adjustable parameters.
The coupling constant $g_{KN \Lambda}$ is very close to its lowest limit~\cite{AS}
within broken SU(3)-symmetry.
This parameter, like several other adjustable ones, is driven by CLAS data.
Actually fitting only the SAPHIR data leads to $g_{KN \Lambda}$=9.70.
Finally, the $\chi^2_{d.o.f}$ for the model $M_1$ is significantly
higher than obtained by fitting only the SAPHIR data. Actually the
integrated $\chi^2$ for the latter data set increases by more than
a factor of 2, i.e. the adjustable parameters are driven by the
CLAS data. However, in going from the model $M_1$ to $M_2$ that
integrated $\chi^2$ stays stable, while the integrated $\chi^2$
for CLAS data increases by roughly 30\%. Moreover, in the
integrated $\chi^2$s for the models $M_1$ and $M_2$, CLAS data
represents roughly 55\% and 48\%, respectively, while SAPHIR data
account for about 45\% and 29\%, respectively. These results
indicate that, within our approach, the SAPHIR data show larger
compatibilities with the polarization data, than does the CLAS
data.

\begin{figure}[ht]
\vspace{30pt}
\begin{center}
\mbox{\epsfig{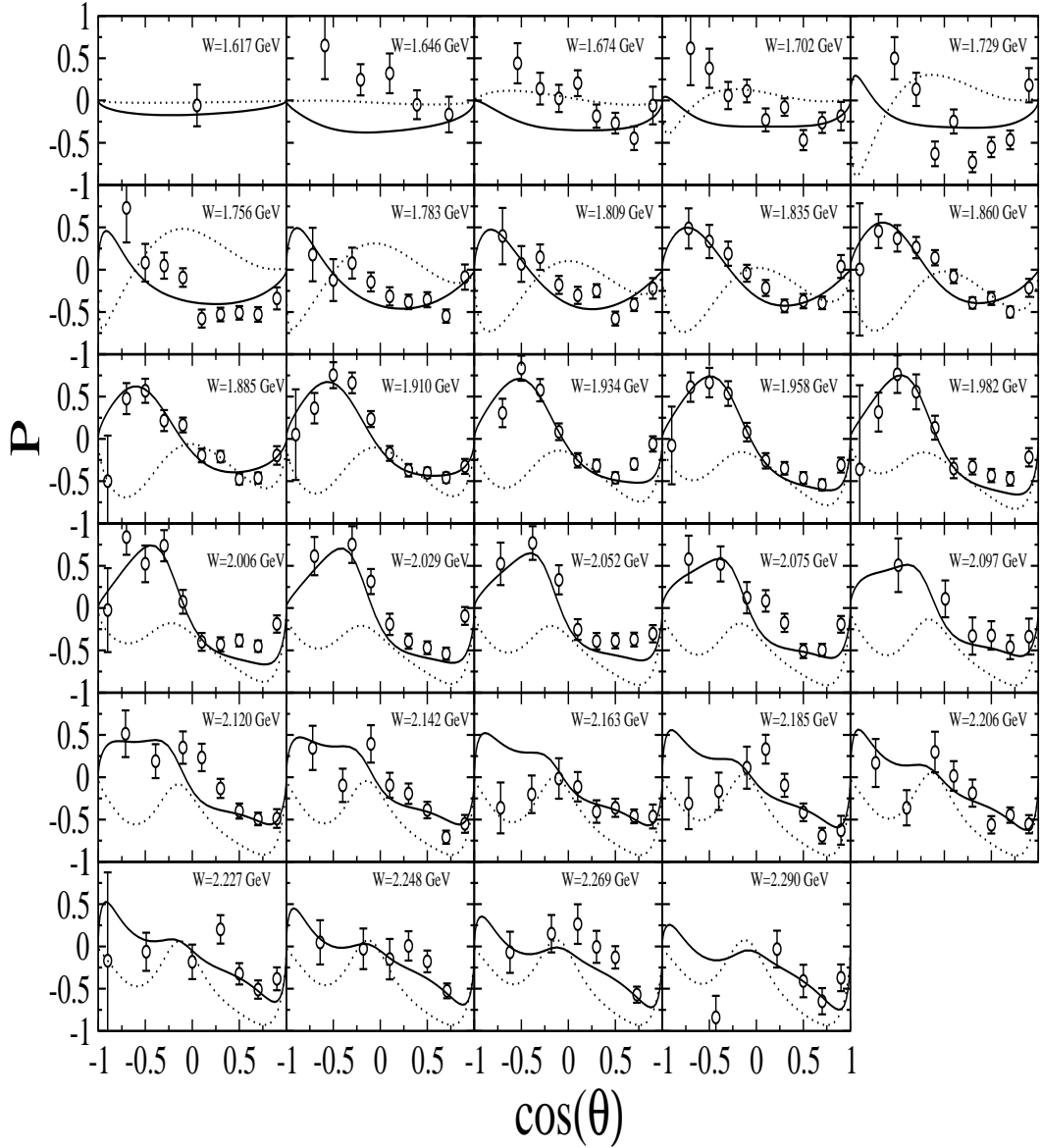}}
\end{center}
\caption{Angular distribution for recoil $\Lambda$ 
polarization asymmetry for the reaction
$\gamma p \to K^+ {\vec \Lambda}$. 
Curves as in Fig.~\ref{fig:dxs-1}.
Data are from Ref.~\cite{JLab04}.}
\protect\label{fig:Recoil-1}
\end{figure}

\begin{figure}[ht]
\vspace{25pt}
\begin{center}
\mbox{\epsfig{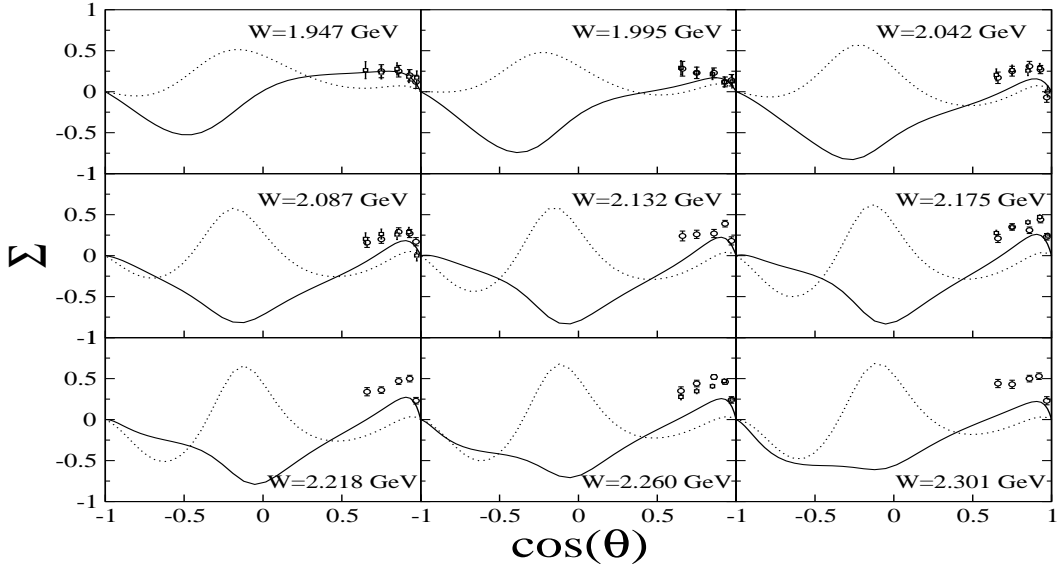}}
\end{center}
\caption{Angular distribution for polarized beam asymmetry for 
the reaction
${\vec \gamma} p \to K^+ \Lambda$. 
Curves as in Fig.~\ref{fig:dxs-1}.
Data are from Ref.~\cite{LEPS} (circles) and Ref.~\cite{LEPS05}
(squares).}
\protect\label{fig:Beam-1}
\end{figure}

\begin{figure}[ht]
\vspace{25pt}
\begin{center}
\mbox{\epsfig{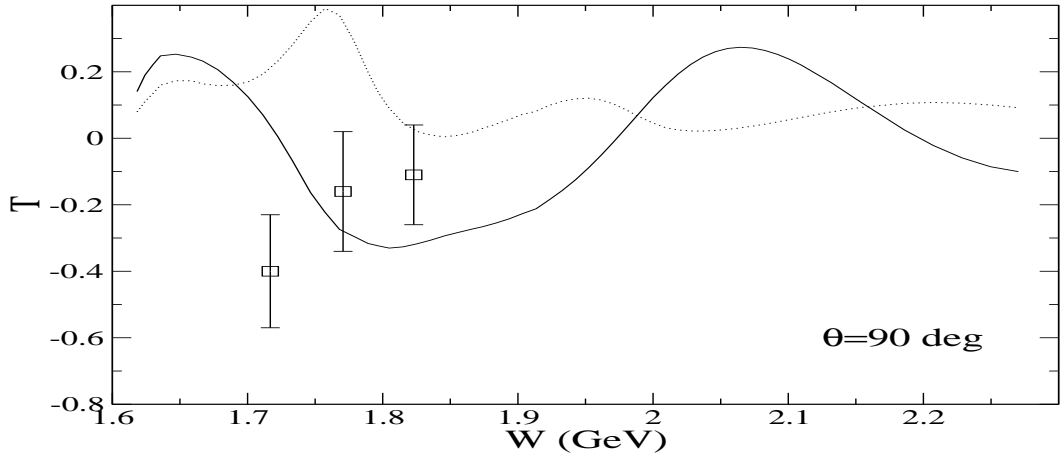}}
\end{center}
\caption{Excitation function for Polarized target asymmetry for 
the reaction
$\gamma {\vec p} \to K^+ \Lambda$. 
Curves as in Fig.~\ref{fig:dxs-1}.
Data are from Ref.~\cite{Target}.}
\protect\label{fig:Target-1}
\end{figure}

In Figs~\ref{fig:dxs-1} to ~\ref{fig:Target-1}, results for models $M_1$ and
$M_2$ are compared with the most recent data.

In Fig.~\ref{fig:dxs-1} excitation functions at 19 angles, for
$\theta _{K}~\approx$ 25$^\circ$ to 150$^\circ$ are shown as a
function of total center-of-mass energy for W~=~1.6 GeV to 2.3
GeV. Except in very few phase space regions, the two models give
identical results. Given the discrepancies between the two fitted
data sets, our models give an acceptable account of the
differential cross-sections. In the same figure, we show also the
very recent LEPS data~\cite{LEPS05} for cos$\theta$ = 0.75 and
0.85. They turn out to be closer to the CLAS data, rather than to
SAPHIR results.

With respect to the polarization observables, we recall that model
$M_1$ ( dotted curve) has been obtained by fitting only the cross
section-data. So, in Figs~\ref{fig:Recoil-1} to
~\ref{fig:Target-1}, dotted curves are predictions. While the full
curves (model $M_2$) result from fits to differential
cross-section {\it and} polarization observables data.

In Fig.~\ref{fig:Recoil-1} angular distribution of polarized
recoil $\Lambda$ asymmetry is depicted for W~$\approx$~1.6 GeV to
2.3 GeV. Models $M_1$ and $M_2$ give significantly different
results and the latter model reproduces the data  quite well,
except for a few lowest energy ones. It is worthwhile mentioning
that although recoil data represents less than 10\% of the data
base points, and contribute to the total $\chi^2$ by the same
percentage, they have a significant effect in the determination of
the model ingredients.

The polarized photon beam asymmetry, Fig.~\ref{fig:Beam-1}, data
stand for less than 2\% of data base points, but generate about
13\% of the total $\chi^2$. We recall that the fitted data come
from Ref.~\cite{LEPS} and are shown as open circles in all 9 cells
of Fig.~\ref{fig:Beam-1}; while the very recent data~\cite{LEPS05},
depicted as open squares, were not included in the fitted data base.
Here, model $M_2$ (solid curves) shows
an improvement over $M_1$(dotted curves) when compared with the
data. According to our results, further measurements of this
observable around $\theta _{K}~\approx$ 90$^\circ$ would put
strong constraints on the models search.

Polarized target asymmetry has been measured only by one
group~\cite{Target} about 3 decades ago. For completeness, we
compare our models with those few data points,
Fig.~\ref{fig:Target-1}, showing that the model $M_2$ gives a
better agreement with those data. Contribution of those data to
the total $\chi^2$ is around 0.1\%.

In summary, the model $M_2$ provides a reasonable description of
the whole data base Figs.~\ref{fig:dxs-1}-\ref{fig:Target-1} and
the comparisons of the resulting
parameters listed in Table III indicate the importance of
having  polarization observables  data in the study of $N^*$
resonances.


\subsubsection{Search for new resonances}
\label{sec:new-res}

For about three decades, several approaches have been predicting
baryon resonances not seen in extensively investigated $\pi N$
channels. issues related to those missing resonances have
recently been reviewed~\cite{CR2000,Bijker,Thoma05}.
 
The search for missing resonances has been initiated by
predictions formulated in three pioneer approaches: i) Relativized
quark formalism~\cite{Cap86,Cap92,Cap94,Cap98}, ii) Algebraic
approach~\cite{Bijker}, iii) Hypercentral constituent quark
model~\cite{Gian}.

Moreover, several authors have reported about three missing resonances
with masses between 1.8 and 2 GeV, namely, $S_{11}$, $P_{13}$, and $D_{13}$.
In the present work, we have investigated possible contributions from
such resonances to the $\gamma p \to K^+ \Lambda$ reaction mechanism.
Before presenting our results, we give a brief account of findings 
reported in the literature with respect to those resonances.
It is worthwhile keeping in mind that
all the results mentioned below and referring to the CLAS data,
use the CLAS 2004 results~\cite{JLab04} and not the more recent
ones~\cite{JLab05}. So, conclusions based on those works have to
be updated. 


{\bf a) Third $S_{11}$}

The extracted values for the mass and width of a new $S_{11}$ are close
to those predicted by the authors of Ref.~\cite{LW96}
(M~=~1.712 GeV and $\Gamma$~=~184 MeV), as a $K Y$ bound state.

The chiral constituent quark approach used in the present work
served~\cite{CQM-1b,LS-3} in the interpretation
of the $\gamma p~\to~\eta p$ data and put forward strong indications
for a third $S_{11}$ with M~=~1.780 GeV and $\Gamma$=280 MeV.

For the one star $S_{11}(2090)$ resonance~\cite{PDG} and where the
mass ranges between 1.880 GeV to 2.180 GeV, the Zagreb group's
coupled channel analysis~\cite{Zagreb} produces the following
values M = 1.792 $\pm$ 0.023 GeV and $\Gamma$ = 360 $\pm$ 49 MeV.
The same one star resonance was invoked in the 1.932 GeV to 1.959
GeV range, using a reggeized isobar model~\cite{WTC} to
investigate the $\gamma p~\to~\eta^\prime p$ reaction. Still
another isobar approach~\cite{Try04} investigation of the $\gamma
p~\to~\eta p$ puts forward an $S_{11}$ resonance with M =1.825 GeV
and $\Gamma$=160 MeV.

A self-consistent analysis of pion scattering and photoproduction
within a coupled channel formalism, indicates~\cite{chen} the
existence of a third $S_{11}$ resonance with M =1.803 $\pm$ 0.007 GeV.

Finally, one of the main recent experimental sources on baryon
resonances comes from  the BES Collaboration~\cite{BES-1,BES-2},
using $J/\Psi$ decay channels. In an early stage,  they
concentrated~\cite{BES-1} on neutral pion and $\eta$ final states:
$J/\Psi \to {\overline{p}}p \pi^{\circ},~{\overline{p}}p \eta$.
The authors could identify the two known $S_{11}$ resonances and
extracted their masses and widths in agreement with the PDG
values. They found a structure at M~=~1800 MeV, the quantum
numbers of which could not be identified because of lack of
statistics.

{\bf b) Third $P_{13}$}

Very recently, the BES Collaboration has released~\cite{BES-2}
data for charged pion final states: $J/\Psi \to
{\overline{p}}\pi^+ n ,~{\overline{n}} \pi^- p$. Besides again
identifying  the two known $S_{11}$ resonances, they put forward
the following interesting results: i) The Roper $P_{11}$
resonance's mass and width are reported, M = 1358 $\pm$ 6 $\pm$ 16
MeV and $\Gamma$ = 179 $\pm$ 26 $\pm$ 50 MeV, to be significantly
smaller than their widely used values. ii) A fourth resonance was
identified by the authors with M = 2068 $\pm$ 3$^{+15}_{-40}$ MeV
and $\Gamma$ = 165 $\pm$ 14 $\pm$ 40 MeV, 3/2$^+$ spin parity.

{\bf c) Third $D_{13}$}

The first indication of a new $D_{13}$ with a mass close to 1.9GeV
was suggested by Mart and Bennhold~\cite{ELA-MB}, who interpreted the
SAPHIR 1998 data~\cite{SAPHIR98} within an isobar approach.
Subsequently, it was shown that those data
could be reproduced both within an isobar model~\cite{Seoul}, 
embodying off-shell effects, and a constituent
quark approach~\cite{CQM-2}. Moreover, recent data~\cite{ELSA}
released in 2004 by the SAPHIR Collaboration did not confirm the
structure reported in their 1998 paper. 
Afterwards, Mart et al.~\cite{Mart04} reached the
conclusion that the manifestations of such a resonance appeared to
be poorly-determined.

Within an isobar model, including {\it s-} and {\it t-}channel
contributions in the tree approximation, Anisovich {\it et
al.}~\cite{Anis05} analyzed the processes $\gamma p~\to~\pi
N,~\eta N, K^+ \Lambda,~K^+ \Sigma^\circ,~K^\circ \Sigma^+$ and
suggested a new $D_{13}$ with M = 1875 $\pm$ 25 and $\Gamma$ = 80
$\pm$ 20. The authors report a less strong indication for an
additional $D_{13}$ with M = 2166 $^{+50} _{-80}$ and $\Gamma$ =
300 $\pm$ 65, that they attribute to the N*(2080) of PDG. However,
recent results from the CB-ELSA Collaboration~\cite{CE-ELSA} on
the $\gamma p~\to~N^*(\Delta ^*)~\to~\pi^\circ p$ puts this latter
two star resonance at M = 1943 $\pm$ 17 and $\Gamma$ = 82 $\pm$
20.

A hybrid isobar plus Regge model has been developed by Corthals et
al.~\cite{REG-Gent}. According to the Regge background model used,
a $D_{13}$(1895) appears or vanishes. The authors suspect a  role
for significant final state interactions not included in their
approach. Such effects are also absent in all isobar models
discussed above.

Such effects, as well as intermediate state  reactions, are of
course embodied in the coupled-channel approaches based on the
K-matrix formalism developed by the Giesssen~\cite{K-matrix-1} and
Groningen~\cite{K-matrix-2} groups, though both groups use isobar
models for the direct processes. Neither of those works show
evidences for new resonances. However, the Giessen group fitted
separately SAPHIR and CLAS 2004 data and the Groningen group used
only SAPHIR data.

Finally, an investigation~\cite{Kelkar} of the relations between
the {\it S}-matrix and time delay in $\pi N$ interactions,
concluded that a $D_{13}$(1940) could appear.

Given the results from other investigations outlined above, we
proceed to the presentation of our findings with respect to 
those new resonances.


%
\begin{table}[ht]
\begin{tabular}{cllllcc}
\hline
\hline
  New resonance && Property &&Model $M_1$&& Model $M_2$ \\
\hline
                && Mass     &&   1.833   &&   1.806  \\
$S_{11} $       && Width    &&   0.288   &&   0.300  \\
                && Strength &&   0.40    &&   0.15  \\[10pt]
                && Mass     &&   1.974   &&   1.893  \\
$P_{13} $       && Width    &&   0.108   &&   0.204  \\
                && Strength &&   0.12    &&   0.28   \\[10pt]
                && Mass     &&   1.912   &&   1.954  \\
$D_{13} $       && Width    &&   0.316   &&   0.249  \\
                && Strength &&   1.50    &&   0.98   \\
\hline
\hline
\end{tabular}
\caption{Determined parameters for the third $S_{11}$, $P_{13} $,
and $D_{13}$ resonances.} \protect\label{tab:cqm-2}
\end{table}

In Section~\ref{sec:param}, we presented our model and made
comparisons with all available data sets. 
In this Section, we use the model $M_{2}$ discussed
in Section~\ref{sec:param}, in order to investigate possible
manifestations of three missing resonances: $S_{11}$, $P_{13}$,
and $D_{13}$. For that purpose, we have attributed 3 adjustable
parameters (mass, width, and strength) to each of those resonances
in the minimization procedure. The extracted parameters are given
in Table~\ref{tab:cqm-2}.

\begin{figure}[ht]
\begin{center}
\mbox{\epsfig{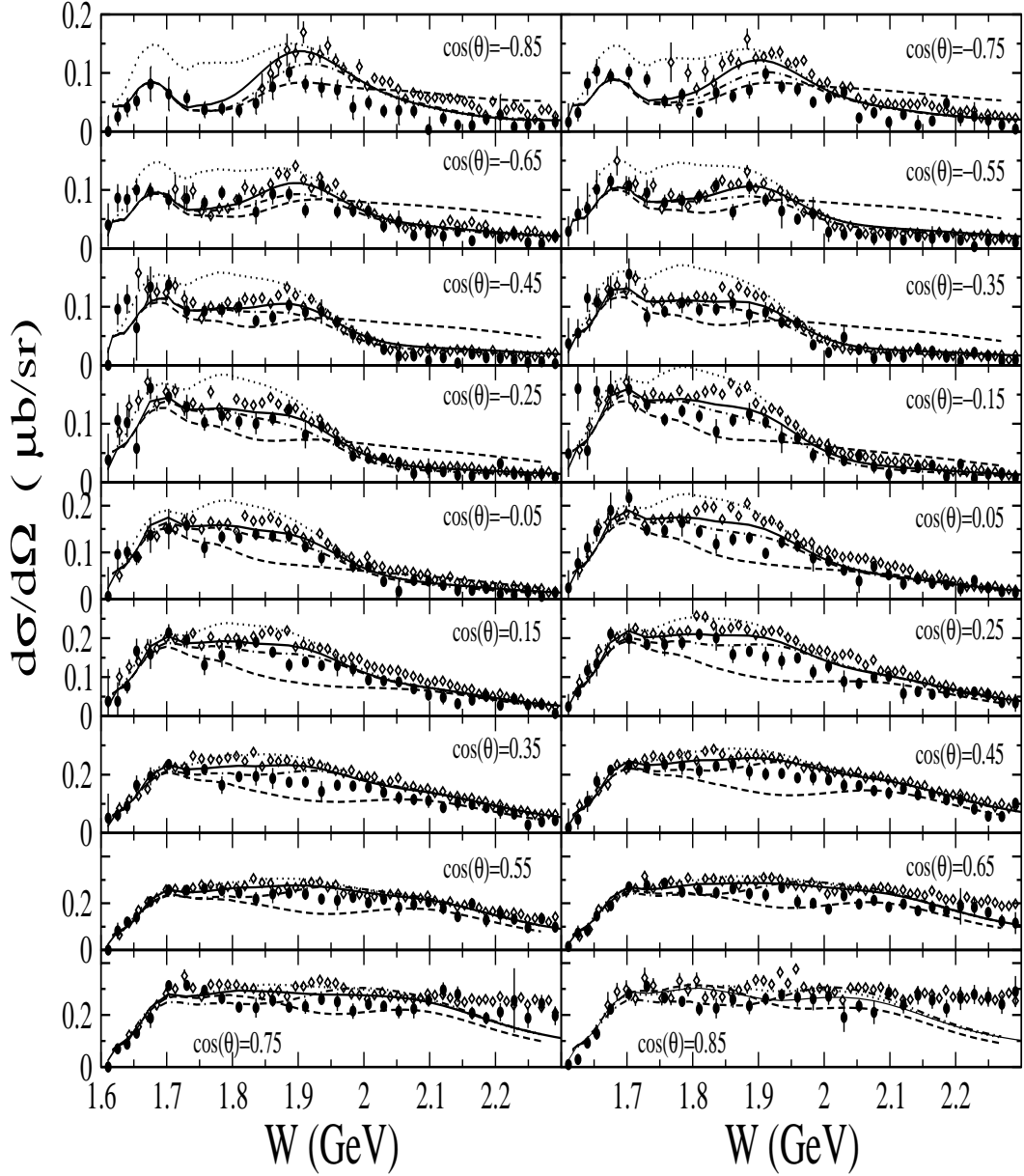}}
\end{center}
\caption{Differential cross-section for the reaction
$\gamma p \to K^+ \Lambda$ as a function of total-center-of-mass energy. 
Solid curve corresponds to the full model $M_2$.
Dotted, dot-dashed, and dashed curves correspond to the full model without
the $3^{rd} S_{11}$, $3^{rd} P_{13}$, and $3^{rd} D_{13}$, respectively.
Data are as in Fig.~\ref{fig:dxs-1}}
\protect\label{fig:dxs-2}
\end{figure}

\begin{figure}[t]
\vspace{25pt}
\begin{center}
\mbox{\epsfig{file=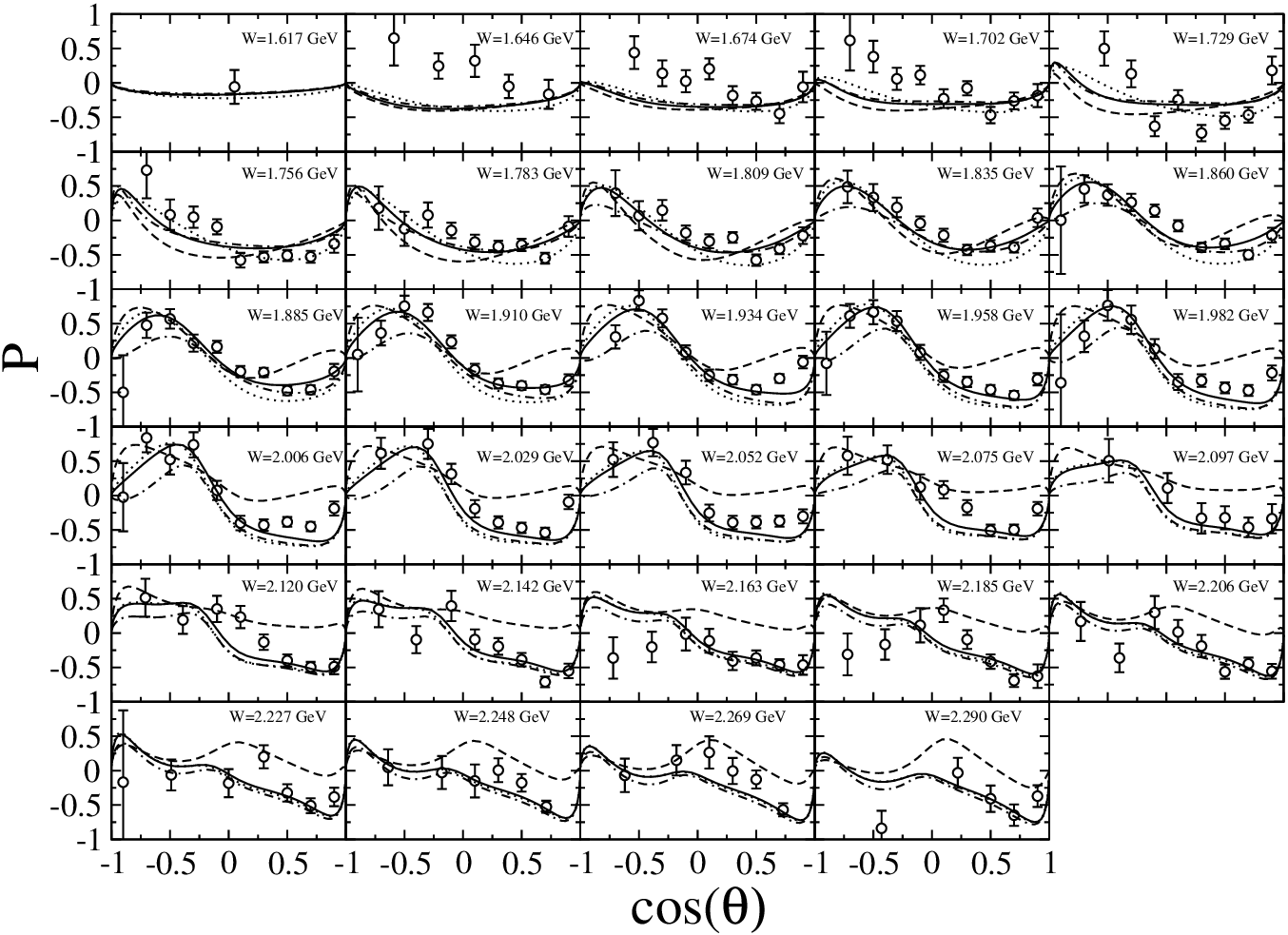, width=140mm, height=167mm}}
\end{center}
\caption{Angular distribution for recoil $\Lambda$ polarization 
asymmetry for the reaction
$\gamma p \to K^+ {\vec \Lambda}$. Curves as in
fig.~\ref{fig:dxs-2}. Data are from Ref.~\cite{JLab04}.}
\protect\label{fig:Recoil-2}
\end{figure}
\begin{figure}[t]
\begin{center}
\mbox{\epsfig{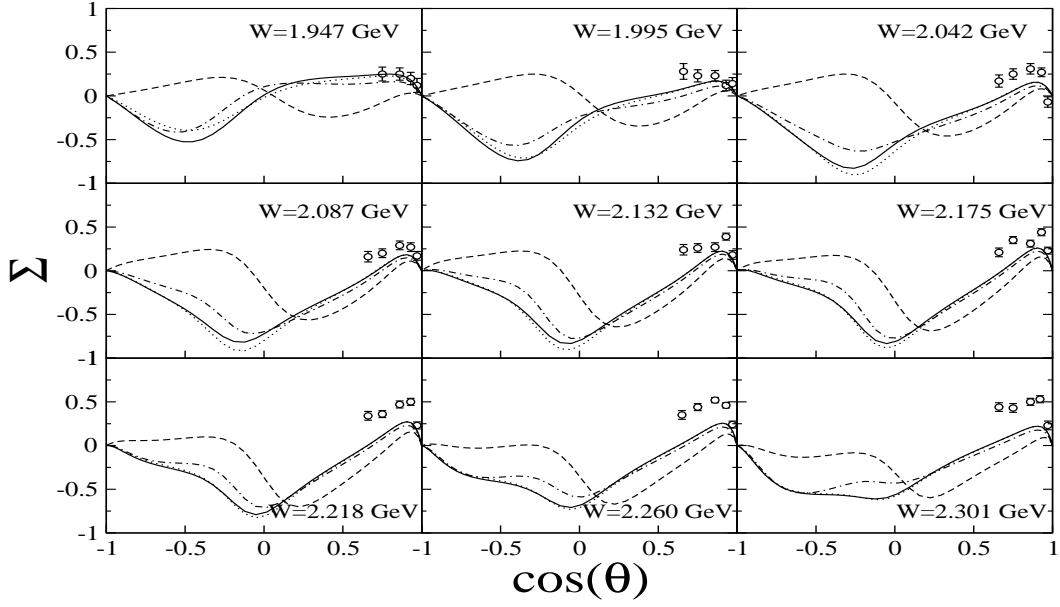}}
\end{center}
\caption{Angular distribution for polarized beam asymmetry for the reaction
${\vec \gamma} p \to K^+ \Lambda$. Curves as in fig.~\ref{fig:dxs-2}.
Data are from Ref.~\cite{LEPS}.}
\protect\label{fig:Beam-2}
\end{figure}
\begin{figure}[hb]
\vspace{25pt}
\begin{center}
\mbox{\epsfig{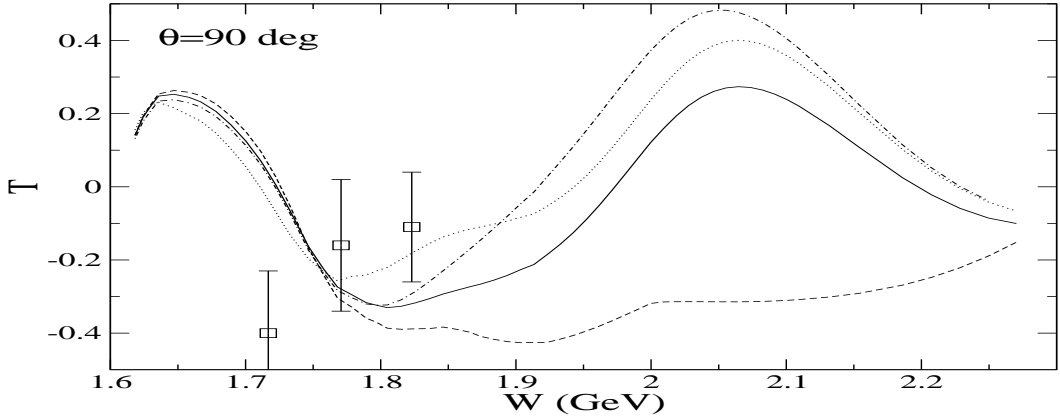}}
\end{center}
\caption{Excitation function for polarized target asymmetry 
for the reaction
$\gamma {\vec p} \to K^+ \Lambda$. Curves as in fig.~\ref{fig:dxs-2}.
Data are from Ref.~\cite{Target}.}
\protect\label{fig:Target-2}
\end{figure}

In order to ascertain the role played by each additional resonance
given in Table~\ref{tab:cqm-2}, we proceed as follows. In
Figs.~\ref{fig:dxs-2} to~\ref{fig:Target-2} we show the same
observables as in Figs.~\ref{fig:dxs-1} to~\ref{fig:Target-1},
respectively. For each observable, the model $M_2$ is depicted
again. The three other curves in Figs.~\ref{fig:dxs-2}
to~\ref{fig:Target-2} correspond to the model $M_2$ with one of
the additional resonances switched off, without further
minimizations. In those figures, the curves are: $M_2$ without the
third $S_{11}$ (dotted curve),
 $P_{13}$ (dot-dashed curve), and $D_{13}$ (dashed curve).

From the differential cross-sections (Fig.~\ref{fig:dxs-2}) we infer that the
3$^{rd}$ $S_{11}$ has a significant role in the backward hemisphere and the
effect gets enhanced in going to most backward angles.
The manifestations of
this resonance vanish for W~$\le$~1.9 GeV. Moreover, the
interference terms due to this resonance appear to be destructive in the
full model $M_2$.

Contributions from the 3$^{rd}$ $P_{13}$ resonance are confined
roughly to the energy range 1.8 $\le$ W $\le$ 2.0 GeV with
increasing magnitude in going from forward to backward angles.
Those contributions are rather small, but non vanishing in the
whole phase space.

The most significant effects due to the  3$^{rd}$ $D_{13}$
resonance are around $\theta _{K}~\approx$ 90$^\circ$ and
W~$\approx$~1.9 GeV. The interference terms come out to be
constructive in the  forward hemisphere in the whole energy range
and in the backward hemisphere for roughly W~$\le$~2.0 GeV.

The recoil hyperon polarization asymmetry,
Fig.~\ref{fig:Recoil-2}, shows no significant sensitivity to the
third $S_{11}$ and $P_{13}$ except in very limited phase space
regions, while switching off the 3$^{rd}$ $D_{13}$ leads to
important variations in the model values for roughly W~$\ge$~1.9
GeV, mainly in the forward hemisphere.

The same trends are observed for the polarized beam asymmetry with respect to the
third $S_{11}$, Fig.~\ref{fig:Beam-2}.
The highest sensitivities to the two other resonances appear in
the backward hemisphere and are significant for the 3$^{rd}$ $D_{13}$.

For the sake of completeness, 
in Fig.~\ref{fig:Target-2} we show
the excitation function at $\theta _{K}$ = 90$^\circ$ for the
polarized beam asymmetry. 
As already mentioned, this observable is
by far the least studied experimentally. Our results might
nevertheless indicate that the 3$^{rd}$ $D_{13}$ produces a
significant structure at higher energies.


%

\subsubsection{Role of resonances in total and differential cross-section,
and polarization observables}
\label{sec:role}
\begin{figure}[hb]
\vspace{30pt}
\begin{center}
\mbox{\epsfig{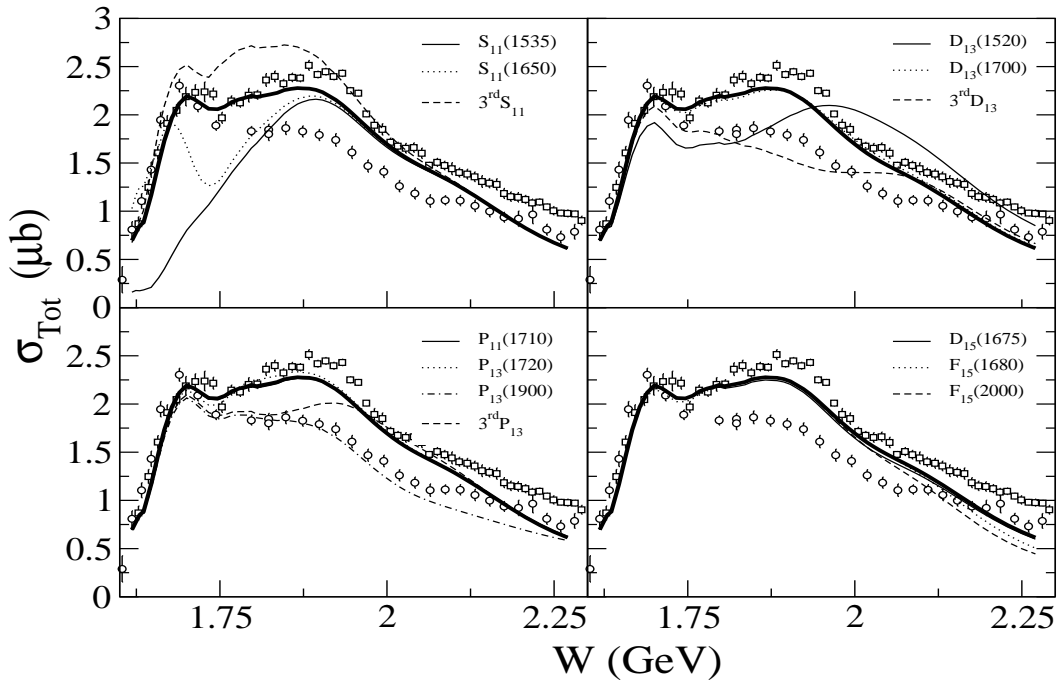}}
\end{center}
\caption{Total cross-section for $\gamma p~\to~K^+ \Lambda$ as a
function of total center-of-mass energy. The bold full curves come
from the model $M_2$. In each cell different curves correspond to
the model $M_2$ with one resonance switched off, as singled out in
each cell. Data are from Refs.~\cite{JLab05,ELSA}.}
\protect\label{fig:tcs}
\end{figure}

Total cross-sections have been extracted by both CLAS and SAPHIR
collaborations. Those data were not included in our fitted data
base. The postdiction of our model $M_2$ is depicted in
Fig.~\ref{fig:tcs} in bold full curves. In each of the four cells,
we show in addition the results of that model $M_2$ with only one
resonance switched off at a time.

The first observation concerns the discrepancies between the two
data sets. As  already pointed out~\cite{CC-05b}, here the
discrepancies are more significant than in the case of
differential cross-sections. This increased discrepancy is likely
due to two facts: i) the two collaborations have performed
measurements in non completely overlapping phase space regions,
ii) different extrapolation methods to the unmeasured angular
areas are used. The total cross-section extracted from
differential cross-sections might then be misleading if included
in the data base and/or used to draw strong conclusions about the
reaction mechanism. There is however a puzzling point, namely,
total cross-sections for the $\gamma p~\to~K^+ \Sigma^\circ$
extracted by the same collaborations, agree quite well with each
other (e.g. see Figs. 20 and 21 in Ref.~\cite{JLab05}).

The model $M_2$ ingredients are dominated by both data sets up to
W~$\approx$~1.7 GeV, by CLAS data up to W~$\approx$~2.0 GeV, and
by SAPHIR data above that region. Two structures appear at about
1.7 GeV and 1.9 GeV.

To gain better insight into the role played by each resonance with
mass M~$\le$~2 GeV, we show curves obtained using the model $M_2$
by switching off each resonance. Table~\ref{tab:Res-1by1} gives
the $\chi^2$ for each case, without further minimizations.

In the following, we concentrate on the model $M_2$ in order to
investigate contributions from various resonances. The points
discussed below do not depend on the total cross-section data, but
they embody effects from all other fitted observables.
Moreover, we present the effects of each resonance with respect to
the fitted data base. Here, in order to limit the number of
figures, we summarize our findings in Table~\ref{tab:Res-1by1},
the content of which is explained below.

The integrated $\chi ^2$ in model $M_2$ can be written as a sum of five partial $\chi ^2_i$s,
\begin{eqnarray}
{\chi ^2} = \sum_{i = 1}^5 {\chi ^2_i}~,
\label{eq:chi}
\end{eqnarray}
where i refers to the data sets, namely,

\noindent i~=~1~: CLAS differential cross-sections, $(d\sigma)_{CLAS}$;

\noindent i~=~2~: SAPHIR differential cross-sections, $(d\sigma)_{SAPHIR}$;

\noindent i~=~3~: CLAS recoil polarization asymmetry, $P$;

\noindent i~=~4~: LEPS polarized beam asymmetry, $\Sigma$;

\noindent i~=~5~: Bonn polarized target asymmetry, $T$.

Then, for each switched off resonance, and without further
minimizations, we obtain the relevant integrated 
$[\chi^2]_{M_2-N^*}$ and partial $[\chi^2_i]_{M_2-N^*}$ for the observable
$i$.  (Here, the  subscript ${M_2-N^*}$ denotes that the
particular resonance $N^*$ has been turned off.) Finally, we
define the following ratio:
\begin{eqnarray}
{\mathcal R_i} = \frac {[\chi ^2_i]_{M_2 - N^*}} {[\chi ^2_i]_{M_2}}~,
\label{eq:Ratio}
\end{eqnarray}
which gives a measure of the role of the relevant $N^*$ with
respect to the observable numbered i. In columns 3 to 7 of
Table~\ref{tab:Res-1by1} following found intervals are reported:

\noindent ~1.0~$<~{\mathcal R_i}~<$~1.5 : *

\noindent ~1.5~$\le~{\mathcal R_i}~<$~1.8 : **

\noindent ~2.0~$<~{\mathcal R_i}~<$~4.4 : ***

\noindent ~6.0~$<~{\mathcal R_i}~<$~8.0 : ****

\noindent 10.0~$<~{\mathcal R_i}~<$~13.4 : *****

Thus more stars indicate a larger role for a particular resonance
on a particular observable $i.$

\begin{table}[th]
\begin{tabular}{crcccccc}
\hline
\hline
 Switched off $N^*$             &$(\chi ^2_{M_2 - N^*})_{d.o.f.}$&
~${\mathcal R}_{(d\sigma)_{CLAS}}$~&
~${\mathcal R}_{(d\sigma)_{SAPHIR}}$~&
~~~${\mathcal R}_{P}$~~~&
~~~${\mathcal R}_{\Sigma}$~~~&
~~~${\mathcal R}_{T}$~~~\\
\hline
$S_{11} $(1535)                 & 10.3~~~~~~&***&***&-&*&*\\
$S_{11} $(1650)                 &  5.7~~~~~~&***&*&*&-&* \\
{\boldmath $S_{11}$}{\bf (1806)}&  6.5~~~~~~&***&***&**&*&- \\
${P_{11}}$(1710)                &  3.3~~~~~~&*&*&*&-&* \\
${P_{13}}$(1720)                &  3.4~~~~~~&-&*&*&*&- \\
${P_{13}}$(1900)                & 13.0~~~~~~&****&-&***&***&- \\
{\boldmath $P_{13}$}{\bf (1893)}&  4.6~~~~~~&**&-&*&*&- \\
${D_{13}}$(1520)                & 20.0~~~~~~&*****&***&***&-&** \\
${D_{13}}$(1700)                &  3.5~~~~~~&*&*&*&-&* \\
{\boldmath $D_{13}$}{\bf (1954)}& 26.4~~~~~~&*****&***&***&***&* \\
${D_{15}}$(1675)                &  3.6~~~~~~&*&*&-&*&* \\
${F_{15}}$(1680)                &  5.0~~~~~~&**&*&**&*&- \\
${F_{15}}$(2000)                &  7.1~~~~~~&***&*&*&***&* \\
\hline
\hline
\end{tabular}
\caption{Schematic presentation of the role played by each resonance in the process
$\gamma p~\to~K^+ \Lambda$.
First column: switched off resonance in model $M_2$; second column: reduced $\chi^2$
without further minimizations to be compared with the $(\chi^2_{M_2})_{d.o.f}$~=~3.3 for the
model $M_2$ (see Table~\ref{tab:cqm-1}). The third to sevenths columns give the
intervals of ${\mathcal R}_i$ (Eq.~\ref{eq:Ratio}) with the number of stars as defined in
the text. The three new resonances investigated here are given in bold.
 }
\protect\label{tab:Res-1by1}
\end{table}


In few cases, the ${\mathcal R_i}$ is slightly smaller than 1.01, shown by a hyphen (-) in that
Table.

The cell on left-top (Fig.~\ref{fig:tcs}) shows the effects of
$S_{11}$ resonances. The lightest resonance affects the total
cross section significantly above its mass, due to constructive
interference terms, and contributes clearly to the first maximum.
This is also the case for the $S_{11}$(1650), with smaller effects
close to threshold. The 3$^{rd}$ $S_{11}$ intervenes around 1.8
GeV and brings in destructive interference. The first and third
$S_{11}$ resonances play important roles
(Table~\ref{tab:Res-1by1}) in the differential cross-section data
from CLAS and SAPHIR, while the second one is present only in the
CLAS data.

For the P-waves (Fig.~\ref{fig:tcs}, left-bottom cell), $P_{11}$(1710) and $P_{13}$(1720)
have negligible
contributions and they do not appear in any of the observables (Table~\ref{tab:Res-1by1}).
According to the same Figure and Table, the $P_{13}$(1900) has strong manifestations within
the CLAS differential cross-sections and, to a less extent, in the $P$ and $T$ polarization
observables.


In the  spin 3/2 D-waves case (Fig.~\ref{fig:tcs}, right-top
cell), the first such state plays an important role with
interference effects turning from constructive to destructive
around 1.9 GeV. Table~\ref{tab:Res-1by1} shows that the
$D_{13}$(1520) is a crucial ingredient in reproducing the CLAS
data and is important with respect to the SAPHIR results. The role
of the $D_{13}$(1700) is negligible, while the 3$^{rd}$ $D_{13}$
has a clear role between roughly 1.8 GeV and 2.0 GeV
(Fig.~\ref{fig:tcs}) and turns out to be a key element,
Table~\ref{tab:Res-1by1}, for all observables, except $T$.

The spin 5/2 D- and F-waves show no significant effects in the total cross-section
(Fig.~\ref{fig:tcs}, right-bottom cell). However, Table~\ref{tab:Res-1by1} underlines
the importance of the $F_{15}$ resonances, especially the second one.

To summarize this Subsection, we find that:

\begin{itemize}
\item{Among the known resonances, the most relevant ones are:
$S_{11}$(1535), $P_{13}$(1900), and $D_{13}$(1520);}
\item {Three other ones are required by data other than those from
SAPHIR: $S_{11}$(1650), $F_{15}$(1680), and $F_{15}$(2000);}
\item {Among the three new resonances, the $D_{13}$(1954)} plays a
crucial role in all observables, except perhaps in the beam
polarization asymmetry. The $S_{11}$(1806) plays an important role
with respect to both differential cross-section data sets, and the
polarized recoil data. The $P_{13}$(1893) has a less strong role
than the two previous resonances and shows up mainly in the CLAS
cross-section data.
\end{itemize}

As mentioned above, all the curves with one resonance removed and 
depicted in Fig.~\ref{fig:tcs} are obtained
without further minimizations.
For the sake of completeness, we shortly report about all possible 
configurations, after minimization,
without new resonances, with one or two of them included.

The reduced $\chi^2$s are given in Table~\ref{tab:cc-dc}. The second column
shows the result with only known resonances, with $\chi^2_{d.o.f}$=5.7. Adding
either a third $S_{11}$ or $P_{13}$ decreases the $\chi^2_{d.o.f}$ by roughly 
18\% (3$^{rd}$ and 4$^{th}$ columns). While a third $D_{13}$ improves the 
$\chi^2_{d.o.f}$ by about 32\%. Columns 6 to 8 show the effects of combinations
of two new resonances. The smallest $\chi^2_{d.o.f}$ is obtained by the
$S_{11}$ and $D_{13}$ pairs. The last columns recalls the result for model $M_2$.
It is interesting to mention that the mass of those resonances stay stable
within 50 MeV through the 7 configurations. These results support our conclusions
above on the a) important role played by the third $D_{13}$, b) improvement
due to an additional $S_{11}$, c) less significant contribution from $P_{13}$. 

\begin{table}[ht]
\begin{tabular}{crccccccc}
\hline
\hline
New resonances $\to$ & ~None~ & ~~${\mathcal S}$~~&
 ~~${\mathcal P}$~~ & ~~${\mathcal D}$~~ & 
 ~~${\mathcal SP}$~~ & ~~${\mathcal PD}$~~ &
 ~~${\mathcal SD}$~~ & ~~${\mathcal SPD}$~~ \\
\hline
$\chi^2_{d.o.f}$ &  5.7~~ & 4.8 & 4.7 & 3.9 & 4.2 & 3.6 & 3.4 & 3.3\\
\hline
\hline
\end{tabular}
\caption{Dependence of the $\chi^2_{d.o.f}$ on all possible combinations 
with respect to the three new resonances.
 }
\protect\label{tab:cc-dc}
\end{table}

Finally, we outline here the results obtained by using only the direct-channel
calculation (with all multi-step processes turned-off). Embodying only the
known resonances leads to $\chi^2_{d.o.f} \approx$12, to be compared to
5.7 in Table~\ref{tab:cc-dc}. The $\chi^2_{d.o.f}$ gets improved by adding new 
resonances and goes down to $\approx$4 when the 3 new resonances are included.
Although the $\chi^2_{d.o.f}$ gets an acceptable value, some of the adjustable 
parameters turn out to be irrealistic. 


%
\subsubsection{ Coupled-channel effects}
\label{sec:ccef}
It is important to illustrate here the differences between the
coupled-channel approach presented here and the often used 
approximations in the literature:  
i) the tree-diagrammodels (direct-channel) neglecting multi-step
phenomena, ii) the coupled-channel K-matrix approaches neglecting 
off-shell effects.

A tree-diagram model can be obtained from the formulation presented
in section~\ref{sec:tf} by turning off all multi-step processes. Namely, the
tree-diagram amplitude is simply
\begin{eqnarray}
T^{tree}_{\gamma N, KY} = v_{\gamma N, KY} +
v^R_{\gamma N, KY} \,,
\label{tree}
\end{eqnarray}
where $v_{\gamma N, KY}$ is the non-resonant amplitude
and $v^R_{\gamma N, KY}$ is the resonant amplitude
calculated from Eq.~(\ref{eq:saghaili}).
\begin{figure}[ht]
\vspace{25pt}
\begin{center}
\mbox{\epsfig{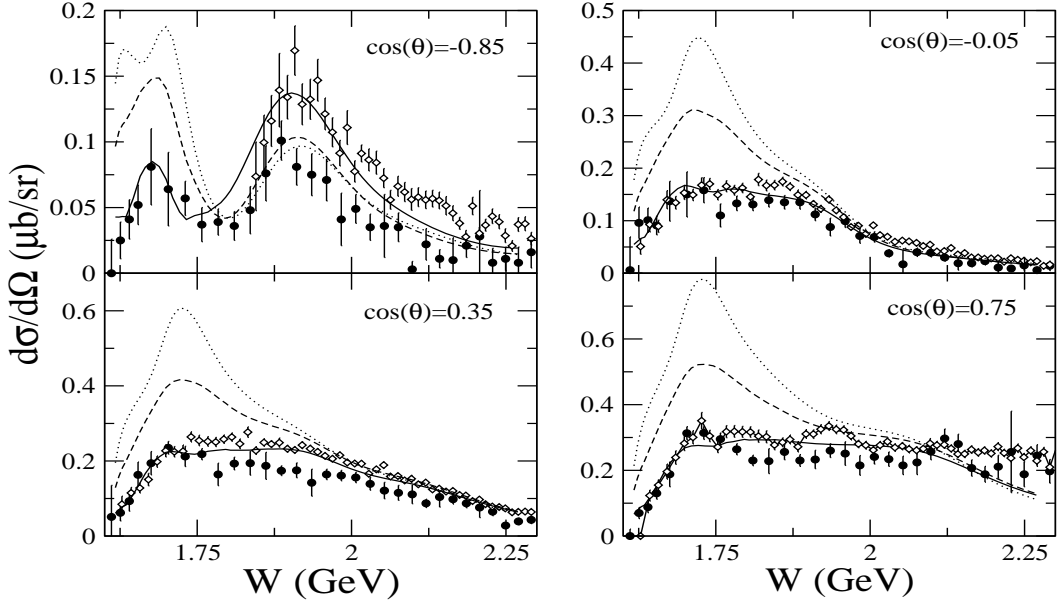}}
\end{center}
\caption{Differential cross-section excitation functions at four 
angles for $\gamma p~\to~ K^+ \Lambda$.
The curves are: model $M_2$ (full curves), direct-channel results
obtained by turning-off multi-step processes in the full calculation
(dotted curves), and off-shell effects swithed-off in the full 
calculation (dashed curves).
Data as in Fig.~\ref{fig:dxs-1}.
} 
\label{fig:cce}
\end{figure}

The importance of the coupled-channel effects can be seen by
comparing the results from Eq.~(\ref{tree}) and the
coupled-channel equations Eqs.~(\ref{gk1a})-(\ref{gk7}). We see in
Fig.~\ref{fig:cce} that when the coupled-channel effects are turned
off, the resulting differential cross sections (dotted  curve)
would largely overestimate the cross sections; especially in the
energy region $W \sim 1.6 - 2 $ GeV. Obviously, the resonance
parameters extracted from using the tree-diagram model will
contain such theoretical uncertainties.

Moreover, within the coupled-channel formalism, the role played
by off-shell effects is depicted in Fig.~\ref{fig:cce}. The
dashed curves there show our results when the off-shell treatment
is turned off. Sizeable effects are present in the same energy
range as above.
\subsubsection{  Meson cloud effects on $N^*$ excitations}
\label{sec:cloud}
In the dynamical study of the $\Delta$ resonance, it was
found~\cite{SatoLee,ky} that the dressed $\gamma N \to \Delta$ transition
contains a large contribution due to the mechanism that the $bare$ $\Delta$
state is not directly excited by the incident photon, but by the pion
first produced by the non-resonant mechanism.
This contribution, commonly termed as the ``meson cloud effect'',
can also be identified within the coupled channel model considered here.
Within the formulation presented in Section II, the meson cloud effect is
contained in the terms within the square brackets of Eq.~(\ref{gk2a}).
Obviously such a meson cloud effect is absent in the tree-diagram model
defined by Eq.~(\ref{tree}).
We also note that the calculation of these meson cloud terms involve
integrations over the off-shell matrix elements of non-resonant amplitudes
$t_{\gamma N, KY}$ and $t_{\gamma N, \pi N}$.
Such off-shell dynamics is neglected in the K-matrix coupled-channel
model~\cite{K-matrix-1}.

\begin{figure}[ht]
\vspace{25pt}
\begin{center}
\mbox{\epsfig{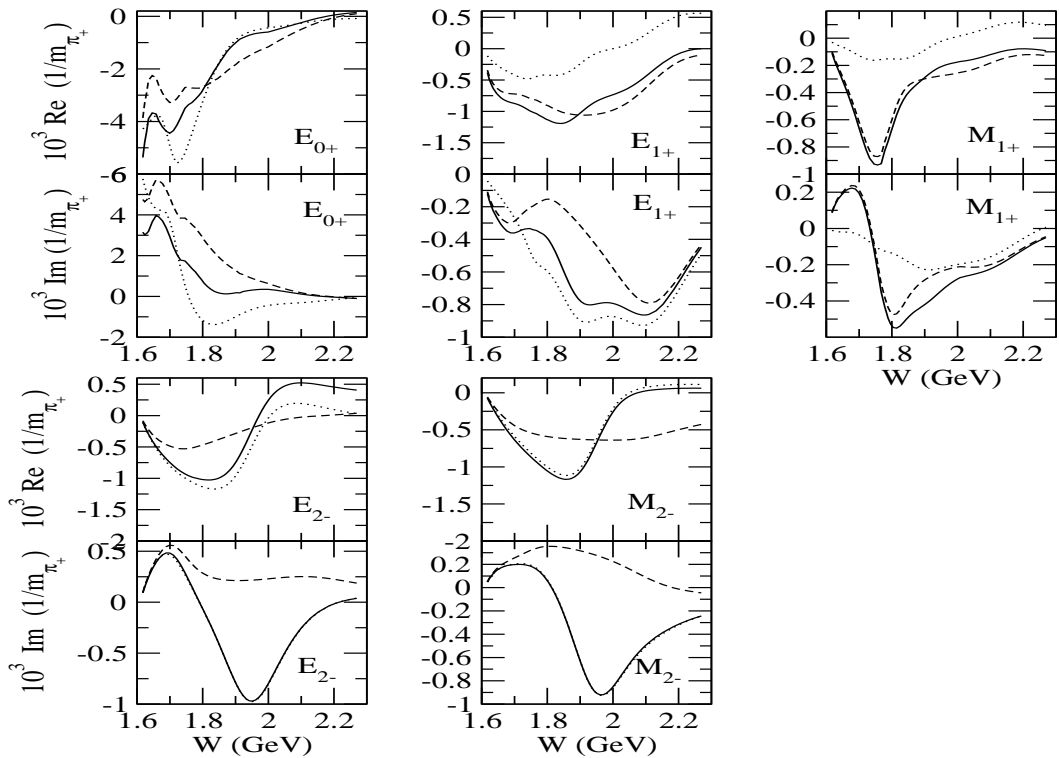}}
\end{center}
\caption{The  multipole amplitudes calculated from the $\gamma N
\to N^* \to KY$ resonant transition for each of the three
considered $S_{11}$-, $P_{13}$-, and $D_{13}$-wave resonances. The
curves are: model $M_2$ (full) and the meson cloud effect
Eq.~(\ref{gk2a}) turned off (dotted). The dashed curves correspond
to the relevant third resonances switched off: $S_{11}$(1806) in
$E_0^+$, $P_{13}$(1893) in $E_1^+$ and $M_1^+$, and $D_{13}$(1954)
in $E_2^-$ and $M_2^-$.} \label{fig:mce}
\end{figure}

The meson cloud effect on the resonances can be illustrated by
comparing the multipole amplitudes calculated with and without the
terms within the square brackets of Eq.~(\ref{gk2a}). Other
quantities of the coupled-channel equations are kept the same in
these two calculations. In Fig.~\ref{fig:mce}, the full curves
correspond to the full $M_2$ model, while the dotted lines are
obtained by turning off terms within the square brackets of
Eq.~(\ref{gk2a}), showing the importance of meson cloud effects in
interpreting the extracted $N^*$ parameters.  To further
understand the meson cloud effects, we need to extend the present
model to investigate electroproduction data such that the $Q^2$
evolution of the multipole amplitude can be extracted, as has been
done in the study of the $\Delta$ resonance of
Ref.~\cite{SatoLee,ky}. Our effort in this direction will be
reported elsewhere.

In that Figure the dashed lines are obtained by switching off the
relevant third resonances investigated here. These results confirm
our conclusions in Sec.\ref{sec:role}, namely, the $D_{13}$(1954)
plays a crucial role, $S_{11}$(1806) has a significant effect, and
contributions from the the $P_{13}$(1893) resonance are smaller
than those from the two other new resonances.


\section{Summary and Conclusions}
\label{sec:Sum}

The main motivation of the present work is the interpretation of
recent associated strangeness photoproduction on the proton, which
require coupled-channel formalisms. In the present work we have
focused on the intermediate state $\pi N$, as well as the
intermediate and final states $KY$ interactions.

We have first applied our formalism to the $\pi p~\to~K Y$ and
$KY~\to~KY$ ($K~\equiv~K^\circ,~K^+$, and
$Y~\equiv~\Lambda,~\Sigma^\circ,~\Sigma^+$) by improving our
previous work~\cite{CC-04} and comparing successfully our results
with the existing data. We have hence fixed the interactions
$v_{\pi N, KY}$ and $v_{KY,KY}$, as well as relevant $N^*$
parameters. Then, starting from the formalism reported in
Ref.~\cite{CC-01}, we have developed a more advanced
coupled-channel approach. For the direct $\gamma p~\to~K^+
\Lambda$ we have used a chiral constituent quark
model~\cite{CQM-1b}. The relevant data have been used to fix the
strengths of intervening resonances within the broken
$SU(6)\otimes O(3)$ symmetry.

Good fits to all of the available data of $\pi^- p \to K^\circ\Sigma^\circ$,
$\pi^- p \to K^\circ\Sigma^\circ$, and $\gamma p \to K^+\Lambda$
have been achieved.
We have demonstrated that the coupled-channel effect can strongly change the
results from the often used tree-diagram models.
We have also found that the meson cloud effects on $\gamma N \to N^*$ are
important in interpreting the extracted resonance parameters.

This work shows that the most relevant known resonances in $\gamma
p~\to~K^+ \Lambda$ process are: $S_{11}$(1535), $P_{13}$(1900),
$D_{13}$(1520), and to a lesser extent $F_{15}$(1680) and
$F_{15}$(2000). Contributions from three new nucleon resonances
have been extensively studied leading to convincing manifestations
of a $D_{13}$ resonance with M~=~1.954 GeV and $\Gamma$~=~249 MeV.
Rather significant effects due to a $S_{11}$ resonance with
M~=~1.806 GeV and $\Gamma$~=~300 MeV is observed. A non negligible
role is also found for a $P_{13}$ resonance with M~=~1.893 GeV and
$\Gamma$~=~204 MeV. Accounts of indications on those resonances
from other sources were summarized.

As a next step, the very new data from LEPS~\cite{LEPS05} and forthcoming
polarized beam data from GRAAL and beam-recoil double polarization
asymmetries from CLAS~\cite{CxCz} and GRAAL, will hopefully clear up 
the experimentalsituation with respect to some inconsistencies within 
the present data base.
Moreover, the ongoing extension of our approach to the
$\gamma p~\to~K^+ \Sigma^\circ,~K^\circ \Sigma^+$ channels will certainly 
bring indeeper insights to the associated strangeness photoproduction 
processes.

Finally, we emphasize that the present coupled-channel calculation
is still far from being complete. While the coupling with the $\pi
N$ channel has been included, it is necessary to extend the
present investigation to include the other channels, in particular
the two-pion channels. Thus the extracted resonance parameters
should be considered preliminary. But they could serve as the
starting points for performing a more advanced coupled-channel
calculation including additional meson-baryon channels (e.g. $\eta
N$, $\omega N$, $\pi\pi N$ ($\sigma N, \pi\Delta, \rho N$)) and to
fit simultaneously all meson photoproduction data.

\begin{acknowledgments}

We are grateful to Wen-Tai Chiang, Zhenping Li, and Tao Ye for
their contributions to earlier stages of this work. We thank
Reinhard A. Schumacher and Mizuki Sumihama for having communicated
to us the CLAS and LEPS data; respectively. The gracious
hospitality during stays in Pittsburgh and visits to Argonne are
very much appreciated by BJD. This work is supported by the U.S.
Department of Energy, Office of Nuclear Physics Division, under
contract N$^o$ W-31-109-ENG-38, and by the US National Science
Foundation under grant PHY-0244526.

\end{acknowledgments}
\newpage
%
%
\begin{appendix}

%

\section{Derivation of scattering equations}  
\label{apdx:Lagrgn}

In this appendix, we show that the scattering Eqs.~(1)-(6) in 
Section~\ref{subsec:cc} can be derived exactly by using the formal 
scattering theory given
in, for example, the text book of Goldberger and Watson~\cite{gold}.
We start with the following Hamiltonian
\begin{eqnarray}
H = H_0 + v + w \,,
\label{eq:apx-1}
\end{eqnarray}
where $H_0$ is the free Hamiltonian, $v$ is the non-resonant 
meson-baryon ($MB$) interaction with $MB = \gamma N, KY, \pi N$, and
\begin{eqnarray}
w=\Gamma^\dagger\frac{1}{E-H_0}\Gamma \,,
\label{eq:apx-2}
\end{eqnarray}
defines the resonant excitation by the $N^*\rightarrow MB$ vertex
interaction $\Gamma$. 

The $MB$ reaction amplitude $T(E)$ is then defined~\cite{gold} by 
(we omit $+ i\epsilon$ in the propagator $1/[E-H_0 + i\epsilon$])
\begin{eqnarray}
T(E) &=& (v+w)[1+\frac{1}{E-H_0}T(E)]  \,,
\label{eq:apx-3}
\end{eqnarray}
or
\begin{eqnarray}
T(E) &=& (v+w) [1 + \frac{1}{E-H_0-v-w}(v+w)]
\label{eq:apx-4} \\
&=& (v+w)
\frac{1}{E-H_0-v-w}(E-H_0) \,.
\label{eq:apx-5}
\end{eqnarray}
Comparing Eqs.~(\ref{eq:apx-3}) and (\ref{eq:apx-5}), we thus have
\begin{eqnarray}
[1+\frac{1}{E-H_0}T(E)] =\frac{1}{E-H_0-v-w}(E-H_0) \,.
\label{eq:apx-6}
\end{eqnarray}

We further define the non-resonant scattering matrix $t$ by
\begin{eqnarray}
t(E)&=&v[1+   \frac{1}{E-H_0}t(E)]
\label{eq:apx-7} \\
&=&v[1+\frac{1}{E-H_0-v}v]
\label{eq:apx-8} \\
&=&[1-v\frac{1}{E-H_0}]^{-1}v 
\label{eq:apx-9} \\
&=&[1+t(E)\frac{1}{E-H_0}]v\,.
\label{eq:apx-9a}
\end{eqnarray}
Eqs.~(\ref{eq:apx-7}) and (\ref{eq:apx-8}) lead to
\begin{eqnarray}
[1+\frac{1}{E-H_0}t(E)] =\frac{1}{E-H_0-v}[E-H_0] \,,
\label{eq:apx-10b}
\end{eqnarray}
and Eqs.~(\ref{eq:apx-9}) and (\ref{eq:apx-9a}) to
\begin{eqnarray}
[1-v\frac{1}{E-H_0}]^{-1} = 1+t(E)\frac{1}{E-H_0}\,.
\label{eq:apx-10a}
\end{eqnarray}

Using Eqs.~(\ref{eq:apx-6}), (\ref{eq:apx-9}), and (\ref{eq:apx-10a}),
 Eq.(\ref{eq:apx-3}) can be written as
\begin{eqnarray}
T(E)&=&[1-v\frac{1}{E-H_0}]^{-1}v
+[1-v\frac{1}{E-H_0}]^{-1}
w[1+\frac{1}{E-H_0}T(E)] \nonumber \\
&=& t(E) + [1+t(E)\frac{1}{E-H_0}]w[1+\frac{1}{E-H_0}T(E)] \nonumber \\
&=& t(E) + [1+t(E)\frac{1}{E-H_0}]w \frac{1}{E-H_0-(v+w)} (E-H_0)\,. 
\label{eq:apx-10}
\end{eqnarray}

We next use the property that
\begin{eqnarray}
\frac{1}{E-H_0-(v+w)}
=[1+\frac{1}{E-H_0-v}t_w]\frac{1}{E-H_0-v}\,,
\label{eq:apx-11}
\end{eqnarray}
with
\begin{eqnarray}
t_w&=& w 
+w\frac{1}{E-H_0-v} t_w\,,
\label{eq:apx-12}
\end{eqnarray}
to write Eq.(\ref{eq:apx-10}) as
\begin{eqnarray}
T(E)&=& t +[1+t(E)\frac{1}{E-H_0}] 
[w+w\frac{1}{E-H_0-v}t_w]\frac{1}{E-H_0-v}[E-H_0] \nonumber \\
&=& t+ 
[1+t(E)\frac{1}{E-H_0}]t_w\frac{1}{E-H_0-v}[E-H_0] \,.
\label{eq:apx-13}
\end{eqnarray}
By using Eq.(\ref{eq:apx-10b}), we then obtain
\begin{eqnarray}
T(E)&=& t +[1+t(E)\frac{1}{E-H_0}]  
t_w[1+\frac{1}{E-H_0}t(E)]\,.
\label{eq:apx-14}
\end{eqnarray}

From the separable form Eq.~(\ref{eq:apx-2}) for $w$, it is easy to find
the solution of Eq.~(\ref{eq:apx-12})
\begin{eqnarray}
t_w=\Gamma^\dagger\frac{1}{E-H_0-\Sigma}\Gamma\,,
\label{eq:apx-15} 
\end{eqnarray}
with
\begin{eqnarray}
\Sigma &=&\Gamma\frac{1}{E-H_0-v}\Gamma^\dagger \\
&=&\bar{\Gamma}\frac{1}{E-H_0}\Gamma^\dagger  \\
&=&{\Gamma}\frac{1}{E-H_0}\bar{\Gamma}^\dagger \,,
\end{eqnarray}
where
\begin{eqnarray}
\bar{\Gamma}^+ = [ 1 + t(E)\frac{1}{E-H_0}]{\Gamma^+}\,, \\
\bar{\Gamma}=\Gamma[1 + \frac{1}{E-H_0} t(E)] \,.
\end{eqnarray}
Substituting Eq.~(\ref{eq:apx-15})  into Eq.~(\ref{eq:apx-14}), we
finally obtain
\begin{eqnarray}
T(E)&=& t +\bar{\Gamma}^+
\frac{1}{E-H_0-\Sigma} \bar{\Gamma}\,.
\label{eq:apx-16}
\end{eqnarray}

Taking the matrix elements of the relevant equations given above
between the channels $a,b,c =\gamma N, \pi N, KY$ and noting that
$H_0 |N^*_i> = M^0_{N^*_i} |N^*_i>$ 
in the center of mass 
frame, $G_{a} = <a|\frac{1}{E-H_0}|a>$, and 
$\Gamma_{N^*_i,a} = <N^*_i|\Gamma|a>$, we then obtain
Eqs.~(1)-(6) in Section~\ref{subsec:cc}.


\end{appendix}

\newpage


%

\end{document}